\documentclass {bmcart}

\usepackage{amsthm,amsmath}
\RequirePackage[numbers]{natbib}
\usepackage[utf8]{inputenc} 
\usepackage{booktabs}
\usepackage{multirow}

\usepackage{graphicx}  

\startlocaldefs
\endlocaldefs

\begin{document}
\begin{frontmatter}

\begin{fmbox}
\dochead{Research}


\title{Safe, efficient and socially-compatible decision of automated vehicles: a case study of unsignalized intersection driving}


\author[
  addressref={aff1},                   
  corref={aff1},                       
  email={dfli@zju.edu.cn}   
]{\inits{DL}\fnm{Daofei} \snm{Li}}
\author[
  addressref={aff1},
  email={ao@zju.edu.cn}
]{\inits{AL}\fnm{Ao} \snm{Liu}}
\author[
  addressref={aff1},
  email={22160514@zju.edu.cn}
]{\inits{HP}\fnm{Hao} \snm{Pan}}
\author[
  addressref={aff1},
  email={cwt525@zju.edu.cn}
]{\inits{WC}\fnm{Wentao} \snm{Chen}}


\address[id=aff1]{
  \orgdiv{Institute of Power Machinery and Vehicular Engineering, Faculty of Engineering},             
  \orgname{Zhejiang University},          
  \street{No. 38 Zheda Road, Xihu District},
  \postcode{310027},
  \city{Hangzhou},                              
  \cny{China}                                    
}



\end{fmbox}


\begin{abstractbox}

\begin{abstract} 
Safe and smooth interacting with other vehicles is one of the ultimate goals of driving automation. However, recent reports of demonstrative deployments of automated vehicles (AVs) indicate that AVs are still difficult to meet the expectation of other interacting drivers, which leads to several AV accidents involving human-driven vehicles (HVs). This is most likely due to the lack of understanding about the dynamic interaction process, especially about the human drivers. By investigating the causes of 4,300 video clips of traffic accidents, we find that the limited dynamic visual field of drivers is one leading factor in inter-vehicle interaction accidents, especially in those involving trucks. A game-theoretic decision algorithm considering social compatibility is proposed to handle the interaction with a human-driven truck at an unsignalized intersection. Starting from a probabilistic model for the visual field characteristics of truck drivers, social fitness and reciprocal altruism in the decision are incorporated in the game payoff design. Human-in-the-loop experiments are carried out, in which 24 subjects are invited to drive and interact with AVs deployed with the proposed algorithm and two comparison algorithms. Totally 207 cases of intersection interactions are obtained and analyzed, which shows that the proposed decision-making algorithm can not only improve both safety and time efficiency, but also make AV decisions more in line with the expectation of interacting human drivers. These findings can help inform the design of automated driving decision algorithms, to ensure that AVs can be safely and efficiently integrated into the human-dominated traffic.

\end{abstract}




\end{abstractbox}
%

\end{frontmatter}




\section{Introduction}\label{Introduction}

Automated driving (AD) is evolving rapidly in recent years. By assisting human drivers in driving tasks, e.g. lane keeping and speed control, AD has achieved much success in  commercialization. Further, the last five years have witnessed the rapid development of autonomous driving technology, e.g. RoboTaxi, which has attracted much attention from both the public and the research community. Although AD is often advertised as a safety and comfort feature in modern vehicles, the current AD technologies have still raised many safety concerns related to human factors \cite{kyriakidis2019}. Without resolving safety concerns and achieving stable driving performances, AD is still far away from winning wide trust from users \cite{manchon2021,Liusa2021,xu_when_2021,yu_measurement_2021}.
As pointed out by Noy et al. \cite{Noy2022}, AD should be designed in accordance to cybernetics principles, i.e. by using a human-centric approach in technology development. 

Such human-centric approach does not mean to consider only the human driver/passenger in AV, but also those traffic participants outside the AV cabin. In the foreseeable future, highly automated vehicles are hopefully to share the open roads with human-driven vehicles (HVs). Considering the infinite varieties of human driving behaviors, it is challenging for AVs to safely and efficiently interact with HVs in dynamic scenarios \cite{li2021planning}. Concerns over the harmonious coexistence of AVs and HVs have been raised by both the academics and industries \cite{Di2021TRC}. Public safety reports indicate that current AVs are driving in unexpected ways from human drivers’ point of view, which leads to several traffic accidents. A road test report by Waymo also shows that the human driver is the critical factor in the interactions between AVs and HVs, posing a significant threat to AVs’ safety \cite{ss2}. However, available driving decision-making algorithms have not sufficiently considered the interactions between AVs and HVs \cite{ss23}. Therefore, there is an urgent need for research on decision-making of automated vehicles in highly dynamic and interaction-intensive driving scenarios.

In current AV decision algorithms, there have been basically two ways to consider inter-vehicle interactions. A common way is to directly imitate the cooperation and interaction behaviors of human drivers. For example, Beaucorps et al. \cite{ss5} obtained some reference speed profiles of specific styles based on human driving data clustering, which were used to achieve human-like driving in complex interactions. Chen et al. \cite{ss6} proposed an imitation learning framework to design the driving policy for complex urban scenarios. Theoretically, given sufficient interaction data of human driving, such models can provide a satisfactory driving policy that considers social compatibility. However, the imitation-based methods are limited by the completeness of dataset, making them difficult to cope with the corner cases not covered.

Another way is to make interactive decisions and planning based on predicting the interacting vehicle’s future behaviors \cite{ss7,ss9,ss8,ss26,ss28,ss29,ss30}. For example, Sezer et al. \cite{ss7} handled the interaction problem by predicting the interacting driver’s intents with uncertainties, while the parameters of the driver behavior model were selected intuitively, and the human decision mechanism was not considered. Menendez-Romero et al. \cite{ss9} proposed a cooperative driving strategy to consider AV’s safety and comfort expectations, and also the conflict vehicle’s efficiency in merging at highway ramps. An intention prediction algorithm is integrated to provide the system with a “courtesy” behavior. Wang \cite{ss8} modeled the interaction at unsignalized intersections using utility functions of safety and efficiency. The algorithm predicts the other vehicle’s driving directions and calculates the optimal speed planning by analyzing the possible collision points. However, the utility settings do not include the characteristics, such as intent and other psychological factors. To summarize, these prediction-based approaches can model how the AV should respond with social compatibility, if the interacting vehicle behaves as predicted.  However, in dynamic scenarios with intense two-way interactions, the interacting vehicle may be influenced by the AV maneuvers and deviate from the predicted motion, which should be further addressed. As pointed out by a recent review on vehicle motion prediction \cite{Karle2022}, to accommodate highly dynamic interactions between ego vehicle and other traffic participants, the coordination between motion prediction and ego-motion planning is one of the major challenges.

The existing literatures have clarified that a clear understanding of other traffic users is key to safe and efficient driving in interaction-intensive scenarios. However, there are only limited studies on socially compatible decision algorithms for AV. Among them, game theory has been often applied to the interactive decision-making involving multiple traffic participants  \cite{linan2020game,jin2020game,cai2021game,chandra2022gameplan,Chenwentao2021, Wangletian2021,Lidaofei2022lgm,LiTTRA2022}. These approaches formulate an AV decision problem in an integrated framework by considering all players simultaneously, with which the game payoff design can also be viewed as one special case of prediction-based methodology. For example, Li et al. \cite{linan2020game} presented a Leader-Follower game-theoretic algorithm for various parametrized intersection scenarios. Wang et al. \cite{Wangletian2021} proposed an integrated prediction and planning framework that allows the AVs to infer the characteristics of other road users. By learning the weights of selfish, altruistic and mediocre driving behaviors, the socially compatible reward is constructed, which optimizes not only AV’s own rewards, but also its courtesy to others. In our previous work \cite{Lidaofei2022lgm}, Prospect Theory is incorporated for the payoff design in an unsignalized intersection game with two or four vehicles.  In \cite{LiTTRA2022}, we developed a level-k game model for the overtaking behaviours on two-lane two-way highways. In summary, game-theoretic decision algorithms has shown promising performances in modelling human-like decisions. However, there is an urgent need on how to develop a design method for decision payoff of drivers, especially in complex and interactive driving contexts. 

Therefore, many crucial questions related to interactive driving need to be answered. For example, what are the key influencing factors of social compatibility that need to be considered in inter-vehicle interactions? How can social compatibility be realized in AV decision? When interacting with HVs, will social compatibility improve the decision performance of AV, e.g. safety and human driver’s acceptance?

To address these challenges, in this research we attempt to incorporate such social compatibility in the AV decision algorithm, with a specific focus on the visual limitation of interacting human drivers. The contribution of this paper is two-fold. 
\begin{enumerate}
  \item A probabilistic model of the truck driver’s visual field is constructed and applied in AV decision design. The model can estimate the probability of AV being observed by the HV driver during the interaction process. To the best of our knowledge, this is the first attempt to consider the visual limitation of interacting HV drivers in an  AV decision algorithm.
  \item A game-theoretic framework is proposed to incorporate social compatibility into AV decision, for which the safety and efficiency improvements over commonly-used algorithms are validated via human-in-the-loop experiments.
\end{enumerate}

The rest of the paper is organized as follows. Section \ref{Method} briefs the research's motivation, constructs the AV visibility model and introduces the socially compatible decision algorithm. Section \ref{hilExps} details the driving simulator experiment design, while the results and discussions are summarized in Section \ref{Results}. Finally, Section \ref{Conclusion} concludes the paper and discusses some potential future work.


\section{Method}\label{Method}
\subsection{Motivation}

As defined by Ladegård \cite{ss3}, social compatibility (SC) is the integration of social fitness and reciprocity, which represents an agent’s responsiveness in social interactions. Similar to daily interpersonal interactions, interactive driving in traffic, as a kind of interaction on wheels, also needs SC in decision-making. The realization of SC in driving decision should be based on perception and prediction of other road users. In other words, an inter-vehicle interaction starts from the perception of each other. Among other things, visual perception plays a key role in the human driver perception, since it provides most of information for further prediction and planning tasks in driving.

\begin{figure}[h!]
  \includegraphics[width=12cm]{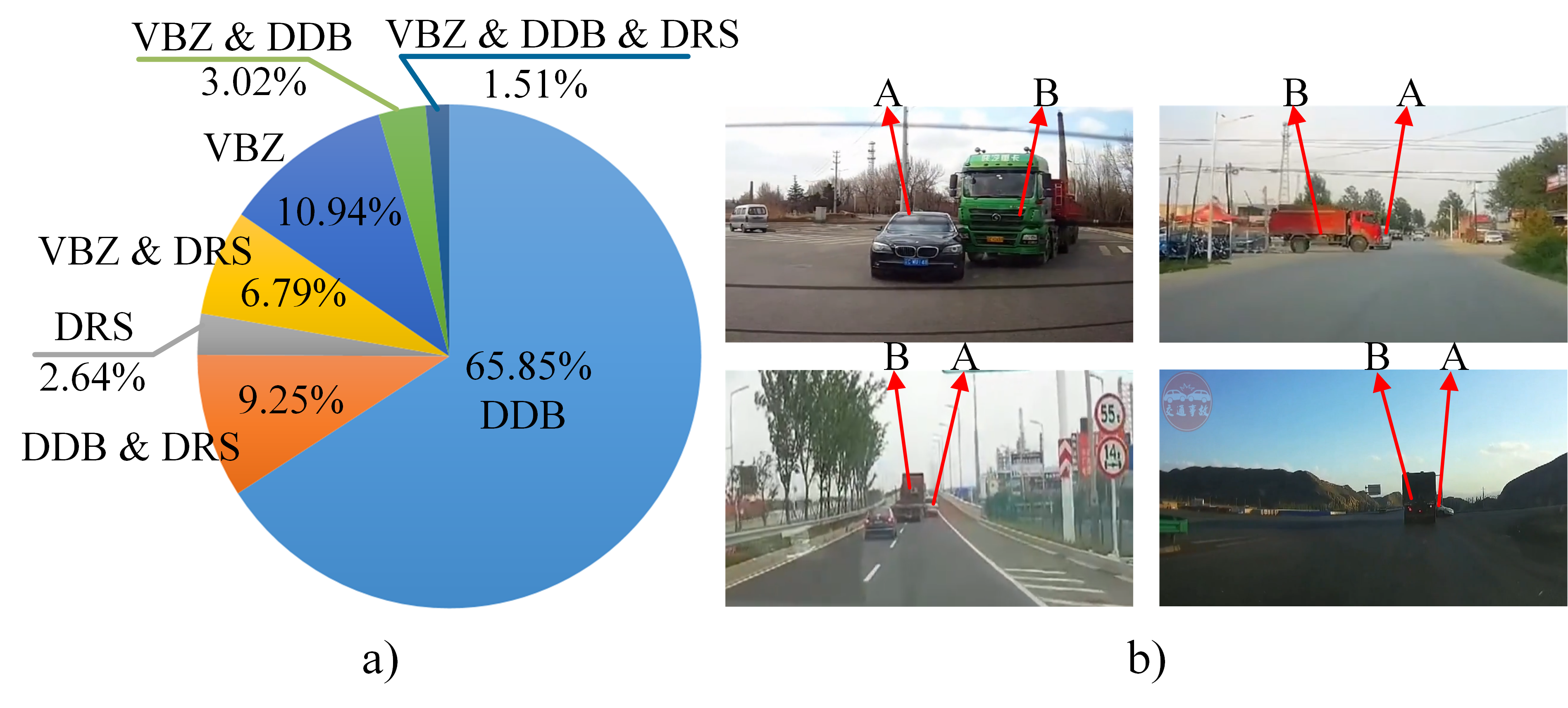}
  \caption{Accident cause statistics in interactive driving. a) Accident causes: Dangerous driving behaviors (DDB); Dangerous road sections (DRS); Visual blind zones (VBZ). b) Accident cases caused by visual blind zones (A: Car; B: Truck). }
  \label{fig1}
\end{figure}

We collected a total of 4,300 video clips of traffic accidents in various scenarios in China (including urban, suburbs, villages and highways), which happened in 4 consecutive months from March 22, 2020 to July 27, 2020 \cite{ss4}. {To investigate the causes of all accidents,} we reviewed the scenarios, including the road/traffic conditions and the vehicle-driver behaviors. The total 530 accidents involving vehicle-vehicle interactions, as shown in Figure \ref{fig1} , were labelled according to their main causes as: (1) dangerous driving behaviors (e.g., emergency braking, crossing multiple lanes in one movement, tailgating), (2) dangerous road sections (e.g., sharp turn, unsignalized intersection, merging ramp), and (3) visual blind zones (e.g., limited view via the rearview mirror, dynamic blind zone due to driver head rotation). 

From Figure \ref{fig1}a, we find that dangerous driving behaviors account for more than 65\% of accidents, however, visual blind zones also contribute to about 22\% of accidents, half of which involve heavy trucks, as shown in the example cases of Figure \ref{fig1}b. It is understandable that for truck drivers, it is more difficult to achieve sufficient searching for visual information, as also pointed out by Larsen \cite{Larsen2004}. Considering the greater severity of truck-involved collisions, safe interaction with truck drivers must be guaranteed in AVs.

When it comes to unsignalized intersections, these visual limitations of human drivers make the interactive driving even more accident-prone. One one hand, in such dynamic driving situations it is challenging for drivers to get accurate perception of the right of way, either of the ego or interacting vehicles. On the other hand, the priority rules to guide vehicle interaction are not clearly predetermined by traffic regulations. In such situations, the aggravated complexities of intersection interactions are safety challenges that AVs have to overcome, especially in those countries and regions that do not strictly enforce the stop or yield sign regulation.

Based on these findings, we believe that the influences of HV driver visual field characteristics should be considered to realize AV's social compatibility in interactive driving. 

\subsection{Framework}
To achieve social compatibility, including social fitness and reciprocity, the AV decision algorithm should (1) promote the HV driver’s understanding of the AV intention, (2) behave with consistency and cooperate tacitly with HV, (3) and consider HV’s interests while guaranteeing AV’s own interests.

\begin{figure}[h!]
  \includegraphics[width=12cm]{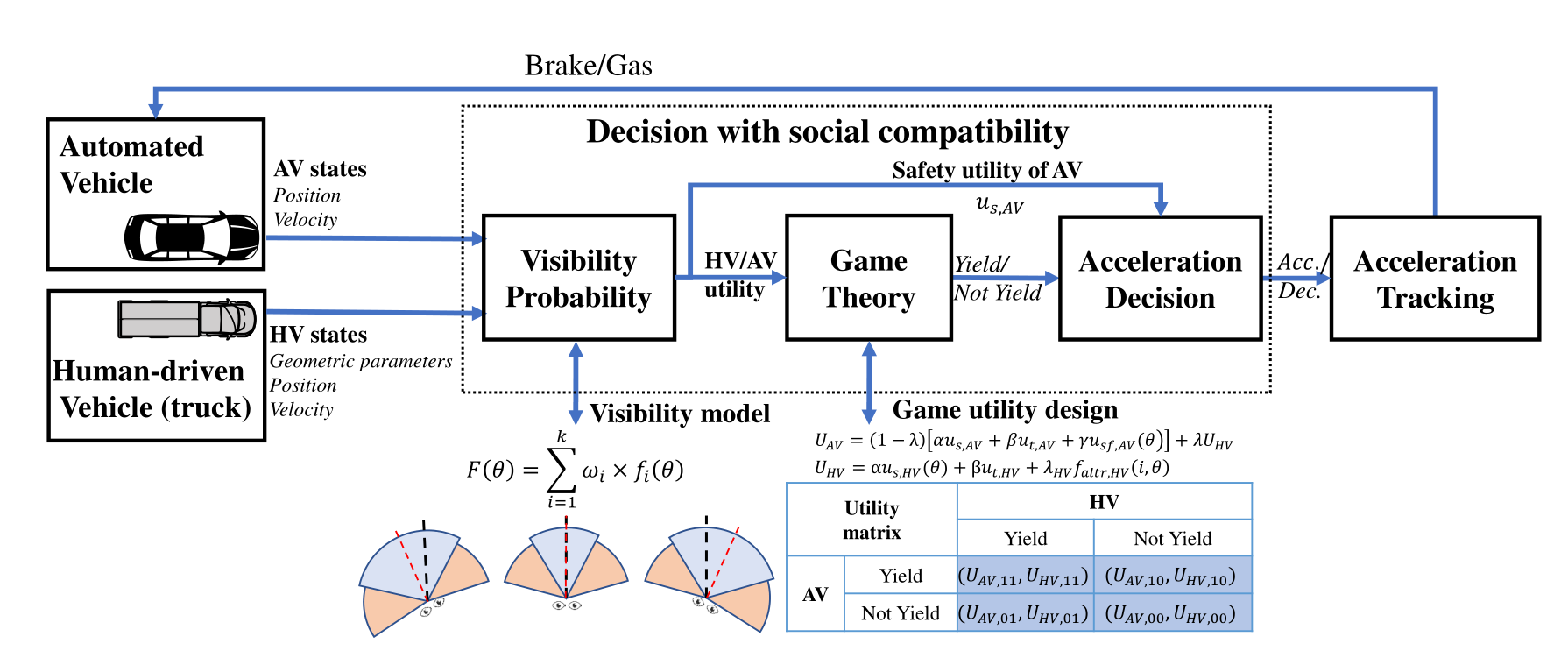}\label{fig2}
  \caption{The overall framework of socially compatible decision algorithm.}

\end{figure}

For the unsignalized intersection scenario, a socially compatible decision framework based on game theory is proposed, as illustrated in Figure \ref{fig2}.

\begin{enumerate}
  \item The inputs of the proposed algorithm consists of two parts, the sensing data of AV states (i.e., position and velocity) and HV states (i.e., geometric parameters, position and velocity).
  \item Based on the sensing data, a visibility probability model is adopted to estimate the truck driver’s visual characteristics, which outputs the probability of AV being observed by the HV driver.
  \item Then, with the designed HV/AV utilitis, the decision game of AV and HV considers safety, efficiency and also social compatibility is solved, which finally outputs a decision of acceleration or deceleration.
  \item Finally, the output decision is executed by the lower level controller.
\end{enumerate}

\subsection{Probabilistic model of AV visibility}
\begin{figure}[h!]
  \includegraphics[width=12cm]{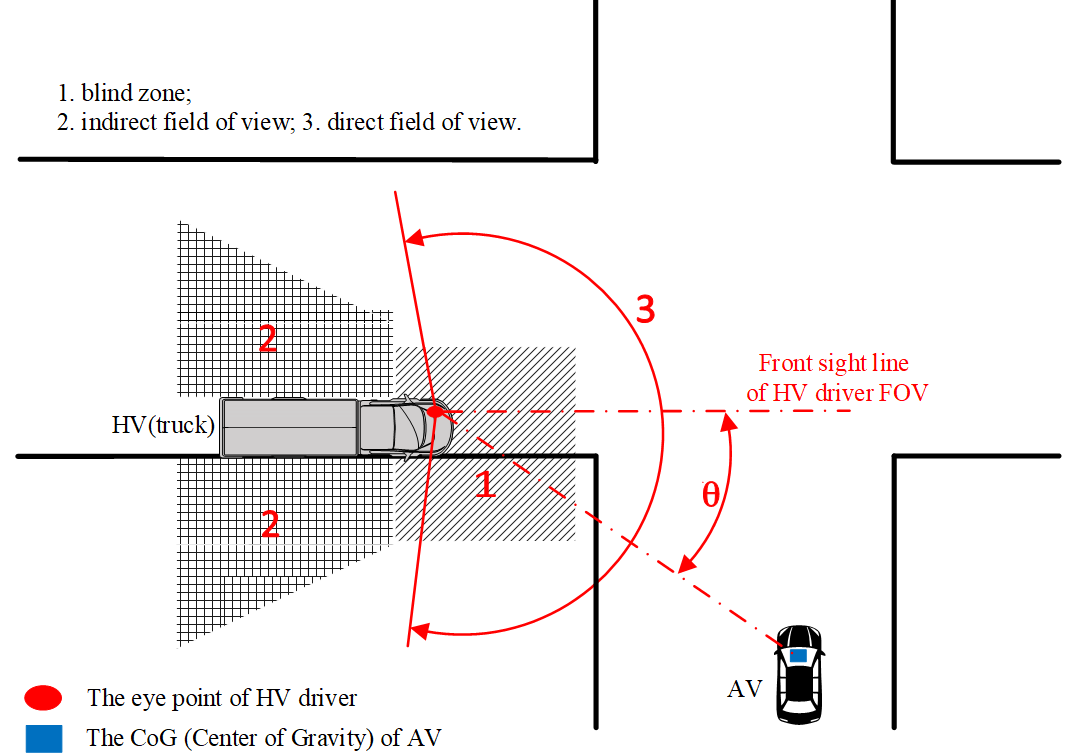}
  
    \caption{Truck driver’s fields of view and blind zones in an unsignalized intersection.}\label{fig3}
  \end{figure}

The two-vehicle interaction in an unsignalized intersection is taken as an example scenario, where the HV is a heavy truck and the ego AV is a passenger vehicle, as schemed in Figure \ref{fig3}. The 360-degree vision of truck drivers can be divided into the blind zones, the direct and indirect fields of view. The direct field of view is the area that can be seen without the aid of any devices. The blind zone is an area around the vehicle that cannot be directly observed when the driver is in a normal sitting position. The indirect field of view can only be seen by using auxiliary devices, e.g. rear-view mirrors. Considering the example intersection, only the blind zone and the direct field of view, i.e. areas 1 and 3, need to be modeled.

\begin{figure}[h!]
  \includegraphics[width=12cm]{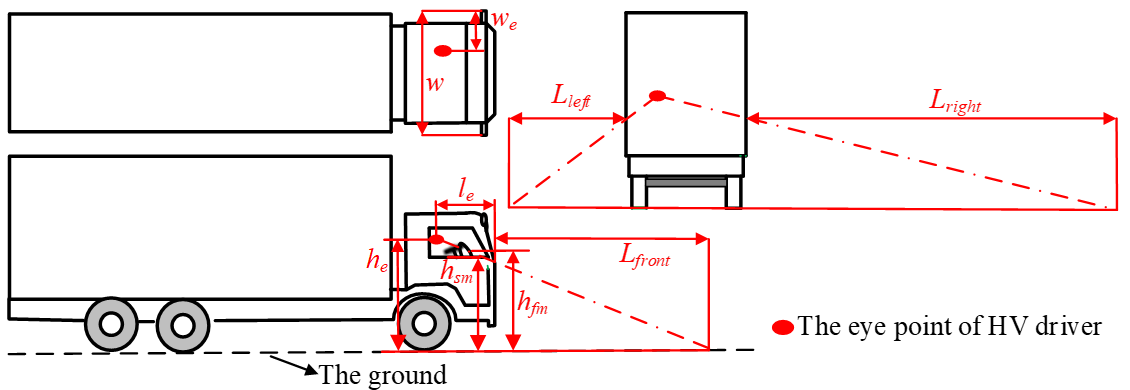}
  
    \caption{The geometric significances of blind zones’ parameters.}\label{fig4}
  \end{figure}

With simplification, the blind zone is defined with 3 parameters, as shown in Figure \ref{fig4}. $L_{left}$ and $L_{right}$ represent the horizontal width of the blind zone on the left/right side of the driver cabin, while $L_{front}$ is its longitudinal length.

\begin{equation}
\label{eq1}
\left\{
      \begin{array}{ll}
        L_{left} &=h_{sm} \times w_e / (h_e - h_{sm}) \\
        L_{right}&=h_{sm} \times (w - w_e) / (h_e - h_{sm}) \\
        L_{front}&=h_{fm} \times l_e / (h_e - h_{fm})

      \end{array}
\right.
\end{equation}
where, $w$ is the overall width of the cockpit, $w_e$ is the distance between the eye point and the left side of the cockpit, and  $l_e$ is the distance between the eye point and the front end of the cockpit. $h_{sm}$ indicates the vertical distance from the bottom edge of the side window to the ground. $h_{fm}$ represents the height of the bottom edge of the windshield/center stack, which blocks the driver’s front line of sight. $h_e$ means the vertical distance from the driver’s eye point to the ground.

\begin{figure}[h!]
  \includegraphics[width=6cm]{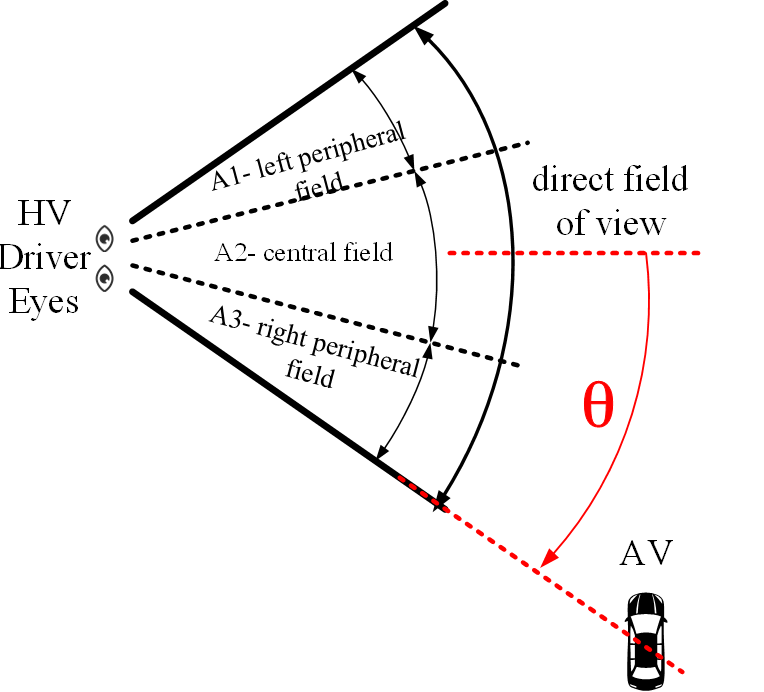}
  
    \caption{The HV driver’s direct field of view.}\label{fig5}
  \end{figure}

For the intersection scenario, {we assume that both AV and HV travel straight, i.e., HV goes from left to right, and AV goes from bottom to top, and only  the AV visibility for areas 1 and 3 in Figure \ref{fig3} are calculated.} When AV is in the blind zone of HV driver, the probability of AV being observed by HV driver is assumed 0. For the front direct field of view (area 3), we further divide it into (A1) the left peripheral, (A2) the central and (A3) the right peripheral sub-fields, as shown in Figure \ref{fig5}.

\begin{figure}[h!]
  \includegraphics[width=12cm]{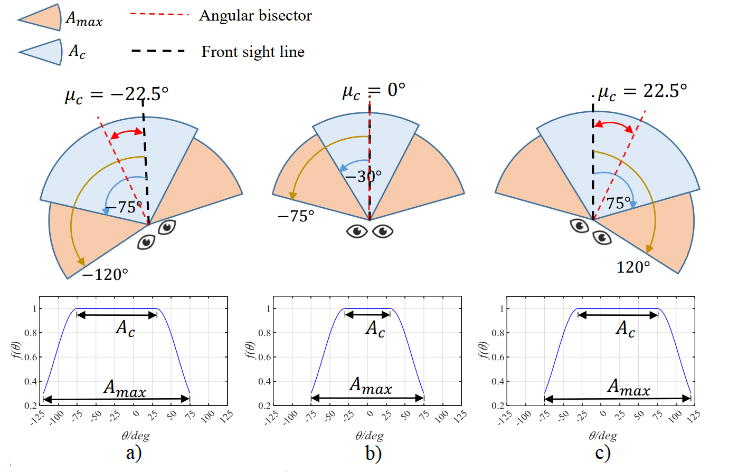}
  \caption{The HV driver visual field and the observation probability function. a) Head turned to the left by $45^{\circ}$. b) Head fixed forward. c) Head turned to the right by $45^{\circ}$.}
    
\label{fig6}
\end{figure}

Assuming that there are normally three natural combinations of head-eye rotation of drivers. (1) If to pay attention to the left, head rotates naturally to the left and eyes rotate freely. (2) If to pay attention to the center, head keeps straight forward and eyes rotate freely. (3) If to pay attention to the right, head rotates naturally to the right and eyes rotate freely. Then the AV’s visibility probability $F(\theta)$ is estimated as follows.

\begin{equation}
\label{eq2}
F(\theta)=\sum_{i=1}^3 \omega_i f_i (\theta),
\end{equation}
where $\theta$ is the viewing angle of AV from the perspective of HV driver. The first part, $\omega_i$, is the probability of HV driver paying attention to the left ($i=1$), center ($i=2$) or right ($i=3$) directions. {The probability $\omega_i$ is ralated to whether there is an object worthy of attention in the specific direction. For the intersection scenario shown in Figure \ref{fig3}, 162 cases from the "DADA-2000" dataset \cite{ss19} are extracted, and then the probabilities $\omega_1$, $\omega_2$, $\omega_3$ are determined according to the statistical results, which are 0, 0.17 and 0.83, respectively.} When HV driver pays attention to the $i$th direction, $f_i(\theta)$ is the observation probability function, representing the probability of AV being observed by the driver, {as shown in Figure \ref{fig6}.}

Human visual observation is affected by the dynamic characteristics of eyeballs. For instance, the macula is located in the optical central area of eye, and its central depression part is the most sensitive area of vision to capture dynamic objects. Therefore, the observation probability function $f_i(\theta)$ in Eq. \ref{eq2} is defined as follows.

\begin{equation}
\label{eq3}
f_i(\theta)=\left\{
      \begin{array}{ll}
        \xi, & \theta \in A_c  \\
        \xi P(\theta), & \theta \in A_{max}-A_c,
      \end{array}
\right.
\end{equation}
where $A_{max}$ is the front direct field’s angular range scanned by driver head rotation, and $A_c$ is the central sub-field’s angular range scanned by head rotation. $\xi$ is a compensation coefficient to consider the environmental factors of the visual capturing ability, e.g. velocity, color and lighting. When AV is in the peripheral sub-fields of view, the AV observation probability of $P(\theta)$ is estimated with the following exponential function.

\begin{equation}
  \label{eq4}
  P(\theta)=p_{min}^{[(2|\theta-\mu_c|-A_c)/(A_{max}-A_c)]^2},
\end{equation}
where $\mu_c$ is the angle between the angular bisector and the front sight line. $p_{min}$ is a minimum visibility probability of AV when it is at the boundary of driver’s peripheral sub-field. If the driver's head is naturally turned to the left or right, an angle of 45 degrees is supposed, then $p_{min}=0.3$, and the HV driver visual field and the observation probability function $f_i(\theta)$ of AV are schemed in Figure \ref{fig6}.

\subsection{Game design considering social compatibility}
The intersection decision game is formulated as a static game, which contains the following elements: the players (AV, HV), the strategy set (Yield, Not Yield) and the utility set ($U_{AV},U_{HV}$). The utility matrix is shown in Table \ref{tab1}, where $(U_{AV,mn},U_{HV,mn})$ is the utility set if AV takes strategy $m$ and HV takes strategy $n$.

\begin{table}[h]
  \caption{The utility matrix of the proposed decision-making algorithm}
  \label{tab1}
  \begin{tabular}{clll}
    \hline
    \multicolumn{2}{c}{\multirow{2}{*}{Utility}} & \multicolumn{2}{c}{HV} \\ 
    \multicolumn{2}{c}{}    &    Yield(1)        &     Not Yield(0)     \\
    \hline
    \multirow{2}{*}{AV}  &  Yield(1)  &  $(U_{AV,11},U_{HV,11})$  & $(U_{AV,11},U_{HV,11})$ \\
                      &  Not Yield(0) &  $(U_{AV,01},U_{HV,01})$  & $(U_{AV,00},U_{HV,00})$ \\
    \hline
  \end{tabular}
\end{table}

\subsubsection*{AV utility}

To achieve safety, traffic efficiency, and also social compatibility, the AV utility $U_{AV}$ is constructed as follows

\begin{equation}
  \label{eq5}
  U_{AV}=(1-\lambda)[\alpha u_{s,AV} + \beta u_{t,AV} + \gamma u_{sf,AV}(\theta)] + \lambda U_{HV},
\end{equation}
where $u_{s,AV}$ and $u_{t,AV}$ are the safety and traffic efficiency utilities of AV, respectively. Social compatibility is represented by both the social fitness utility function $u_{sf,AV}$ and the reciprocal utility, i.e. the HV utility $U_{HV}$. {$\alpha$, $\beta$, $\gamma$, $\lambda$ are the corresponding weights to trade-off among utilities. The AV position variable $\theta$ is used to consider the AV visibility, as shown in Figure \ref{fig5}.}

\subsubsection*{HV utility}

Considering safety, traffic efficiency and reciprocal behavior, the HV utility $U_{HV}$ is designed as

\begin{equation}
  \label{eq6}
  U_{HV}=\alpha u_{s,HV}(\theta) + \beta u_{t,HV} + \lambda_{HV} u_{altr,HV}(\theta).
\end{equation}
where $u_{sf,HV}$ and $u_{t,HV}$ are the safety and traffic efficiency utilities of HV, respectively. $u_{altr,HV}$ is the reciprocal utility of its altruistic behavior, which is weighted by $\lambda_{HV}$.When HV driver yields to AV, {the value of $\lambda_{HV}$ is equal to $\lambda$ in Eq. \ref{eq5}. If the driver does not give way to AV, $\lambda_{HV}$ is 0. The utility functions of AV and HV are further explained in Appendix A. The calibration of parameters, e.g. weighting factors and thresholds, is achieved via randomly-sampled simulations.}

\begin{figure}[h!]
  \includegraphics[width=6cm]{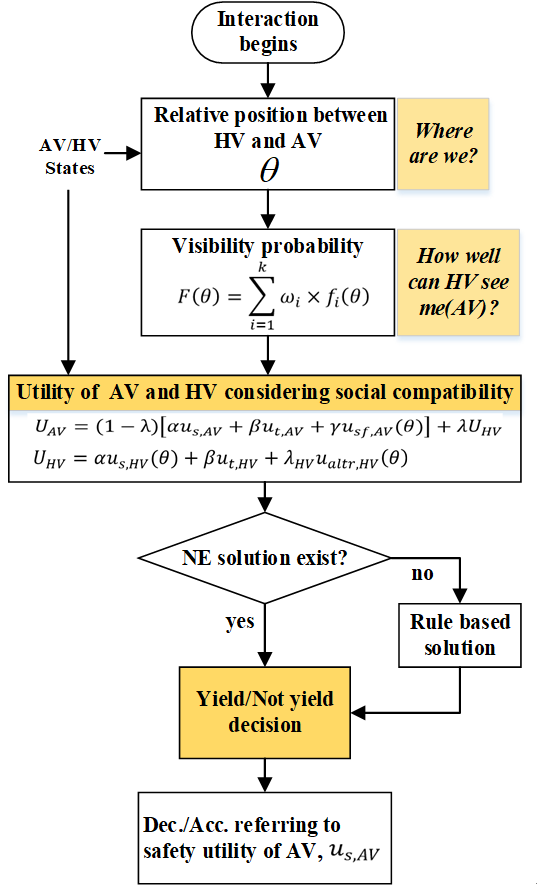}

  \caption{Flowchart of the socially compatible decision algorithm.}\label{fig7}
\end{figure}

To summarize, the flowchart of our socially compatible decision algorithm is presented in Figure \ref{fig7}. Firstly, {the relative position between vehicles is obtained, and is used to calculate the visibility probability of AV. Then, the game utilities are calculated and used to find the Nash Equilibrium (NE) solution. Finally, the specific acceleration/deceleration of AV is decided by combining the yield decision and the safety utility of AV.}

\section{Human-in-the-loop experiments}\label{hilExps}
\subsection{Benchmark algorithms}
Two benchmark decision algorithms are selected to compare with the proposed socially compatible (SC) algorithm. One is the game-based algorithm that only considers safety and traffic efficiency (noSC algorithm), i.e. $\gamma=\lambda=\lambda_{HV}=0$. The other benchmark algorithm is Responsibility Sensitive Strategy (RSS) by Intel Mobileye, which defines a set of safety rules to guarantee "it won't lead to accidents of the autonomous vehicle's blame" \cite{ss12}. RSS is also one of the most popular algorithms that are currently adopted in the academia and industry. The adopted RSS model parameters are listed in Table \ref{tab2}. {The parameter $a_{brake,min}^{HV}$ is determined according to the results of natural driving study in China \cite{ss13}.}

\begin{table}[h!]
  \caption{The parameters in RSS decision algorithm}
  \label{tab2}
    \begin{tabular}{lll}
      \hline
      Parameter  &   Definition    &    Value   \\
      \hline
      $\rho_{AV}$  &  Response time for AV  & $0.5s$ \\
      $\rho_{HV}$  &  Response time for HV  & $2s$ \\
      $a_{accel,max}^{AV}$  &  Maximum acceleration for AV  & $3.5m/s^2$ \\
      $a_{accel,max}^{HV}$  &  Maximum acceleration for HV  & $3m/s^2$ \\
      $a_{brake,min}^{AV}$  &  Minimum deceleration for AV  & $-3m/s^2$ \\
      $a_{brake,min}^{HV}$  &  Minimum deceleration for HV  & $-4.43m/s^2$ \\
      $a_{brake,max}^{AV}$  &  Maximum deceleration for AV  & $-5m/s^2$ \\
      $a_{brake,max}^{HV}$  &  Maximum deceleration for HV  & $-8m/s^2$ \\
      \hline
    \end{tabular}
  \end{table}

\subsection{Apparatus}
\begin{figure}[h!]
  \includegraphics[width=12cm]{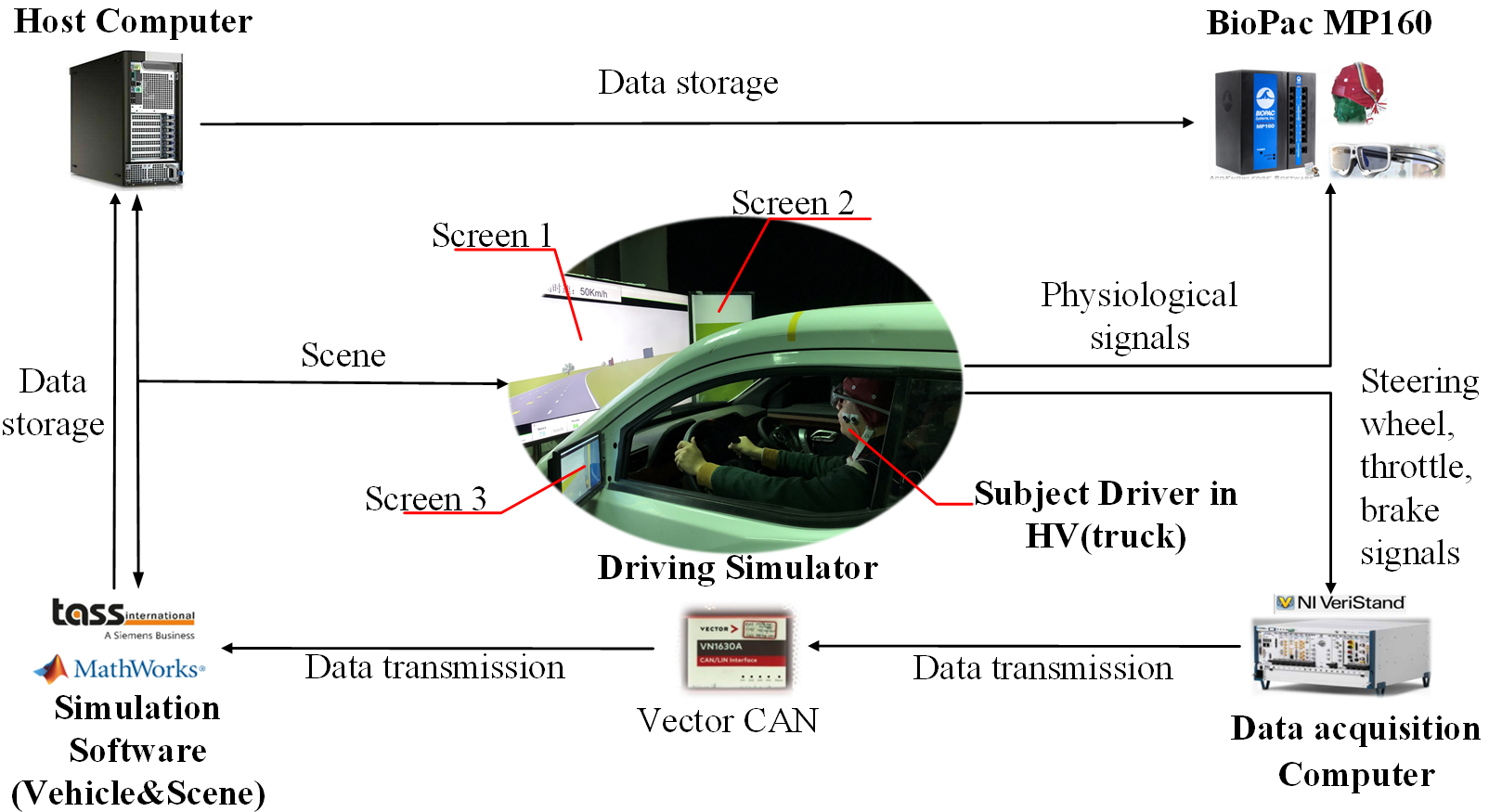}

  \caption{Driving simulator and data acquisition.}\label{fig8}
\end{figure}

As shown in Figure \ref{fig8}, a driving simulator with six degrees of freedom is used as the human-driven truck (HV). The simulator cabin is modified to better reproduce the driver visual limitations in the real truck cabin. The real-time simulation is based on MATLAB and TASS PreScan. The human drivers’ inputs in simulator cabin, i.e. steering, throttle and brake, are collected for the vehicle dynamic model in PreScan, {while the AV algorithm in MATLAB outputs the interaction decisions.} The data of subject Electroencephalogram (EEG) at Fz and Cz positions are recorded and analyzed with BioPac MP160.

\subsection{Participants and experiment design}

We recruited 24 subjects of age between 21 and 28, including 22 males and 2 females. They were asked to drive as in daily driving and to interact with AVs deployed with 3 different decision-making algorithms, namely, 1) noSC algorithm, 2) RSS algorithm, 3) SC algorithm. Three different speed limits were specified, i.e. $20 km/h$ (Lowspd), $45 km/h$ (Midspd), $70 km/h$ (Highspd), respectively. For each algorithm, subjects were asked to drive under specific speed limits in the rightmost lane and to complete 9 HV-AV interactions. When the HV truck was 120 meters away from the conflict area, the AV started with the \emph{same} speed of the truck, to simulate the intense levels of interaction conflict. Once the truck was 100 meters away from the conflict area, the decision algorithm was triggered ON. After each intersection, the HV stopped at the parking area and the subject filled the questionnaire to evaluate the last AV-HV interaction, as detailed in Appendix I.B. Figure \ref{fig9} presents an example of intersection scenario in the experiments.

\begin{figure}[h!]
  \includegraphics[width=8cm]{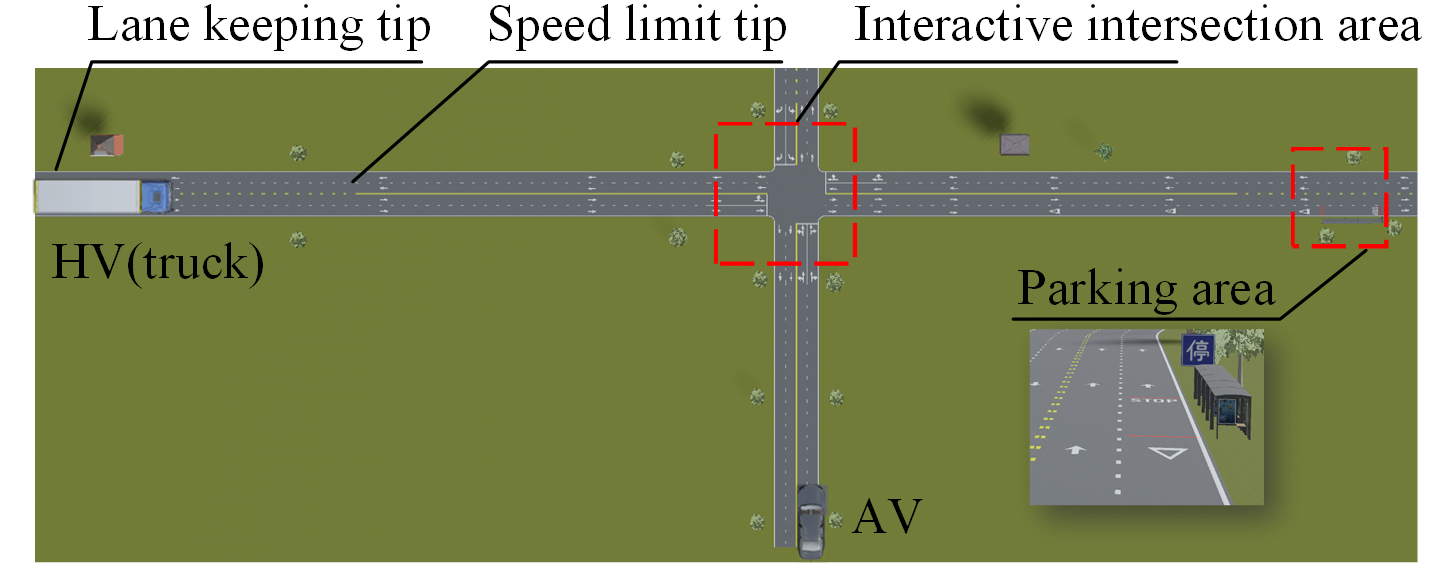}
  \caption{HV-AV interaction in intersection scenes. HV travels from left to right with lane keeping tip and speed limit tip, while AV travels from bottom to top. The HV driver needs to pass through the interactive intersection area safely, and stop in the parking area to finish the questionnaire.}\label{fig9}

\end{figure}

Considering that the physiological data may have a large fluctuation during the interaction and need time to return to a stable state \cite{ss14}, the subjects used $3\sim5$ minutes for free driving before the next interaction. 

An experiment for each subject HV driver took about 90 minutes. The experimental procedure is as follows.
\begin{enumerate}
\item Subject fills in the driver self-ability \cite{ss15} and driving style assessment questionnaires \cite{ss16,ss17}.
\item Subject wears the physiological acquisition devices and confirms the signal recordings.
\item Subject gets familiar with the simulator driving without interaction with AVs.
\item	The formal experiment begins, subject conducts the Lowspd experiment. After each interaction, a subjective questionnaire of driving tasks is filled.
\item	Subject completes \emph{Midspd} and \emph{Highspd} experiments as step 4.
\item	Subject finishes the experiments and takes off the physiological acquisition devices.
\end{enumerate}

\section{Results and discussion}
\label{Results}

Totally 216 interaction cases are obtained, including 207 effective interactions without collisions and 9 failed interactions due to HV’s severe overspeed behaviors (more than $15km/h$ over limit). For detailed analysis, we further divide the interactive cases into 4 speed intervals according to the initial speed triggered by the algorithm, i.e. \emph{Low} ($10-30km/h$), \emph{LowMid} ($30-40km/h$), \emph{Mid} ($40-50km/h$) and \emph{High} ($50-70km/h$). Note that the extreme interaction cases with \emph{High} initial speeds are rare but still possible in real traffic scenarios, which brings severe time pressure to both human drivers and AV algorithms. 

\subsection{Statistical analysis of safety and efficiency}
\label{section_statistics}

{To focus on the intersection interations, we assume that the inter-vehicle interaction ends when one of the vehicles reaches the conflict area}, while the following-up behaviors are not further considered. Therefore, the Time to Arrive (TTA) is selected as the safety evaluation index. When the leading vehicle, either AV or HV, arrives at the conflict area at time $t$, and the lagging vehicle with a speed $v$ is still $L$ distance away from the conflict area, then TTA=$L/v$. If TTA is large, it means when the leading vehicle arrives at the intersection, the lagging vehicle is still far away, so safety can be guaranteed. However, if TTA is too large, the traffic efficiency is compromised since the lagging vehicle is too conservative to make use of the cleared intersection space. Note that if the lagging vehicle fully stops to show its courtesy, a special value TTA=$-1$ is given rather than infinity, and such case is tagged as 'full-stop'. On the other hand, a small TTA means that both vehicles cross the intersection at a very close moment. If TTA is less than a specified threshold, for safety the AV decision algorithm will be overridden by automated emergency braking (AEB) \cite{ss13}. Such case is defined as a \emph{danger} case. Considering the extreme inter-vehicle interactions with high initial speeds, the goal is to minimize the number of danger cases, if not possible to completely avoid all danger cases.

\begin{figure}[h!]
  \includegraphics[width=12cm]{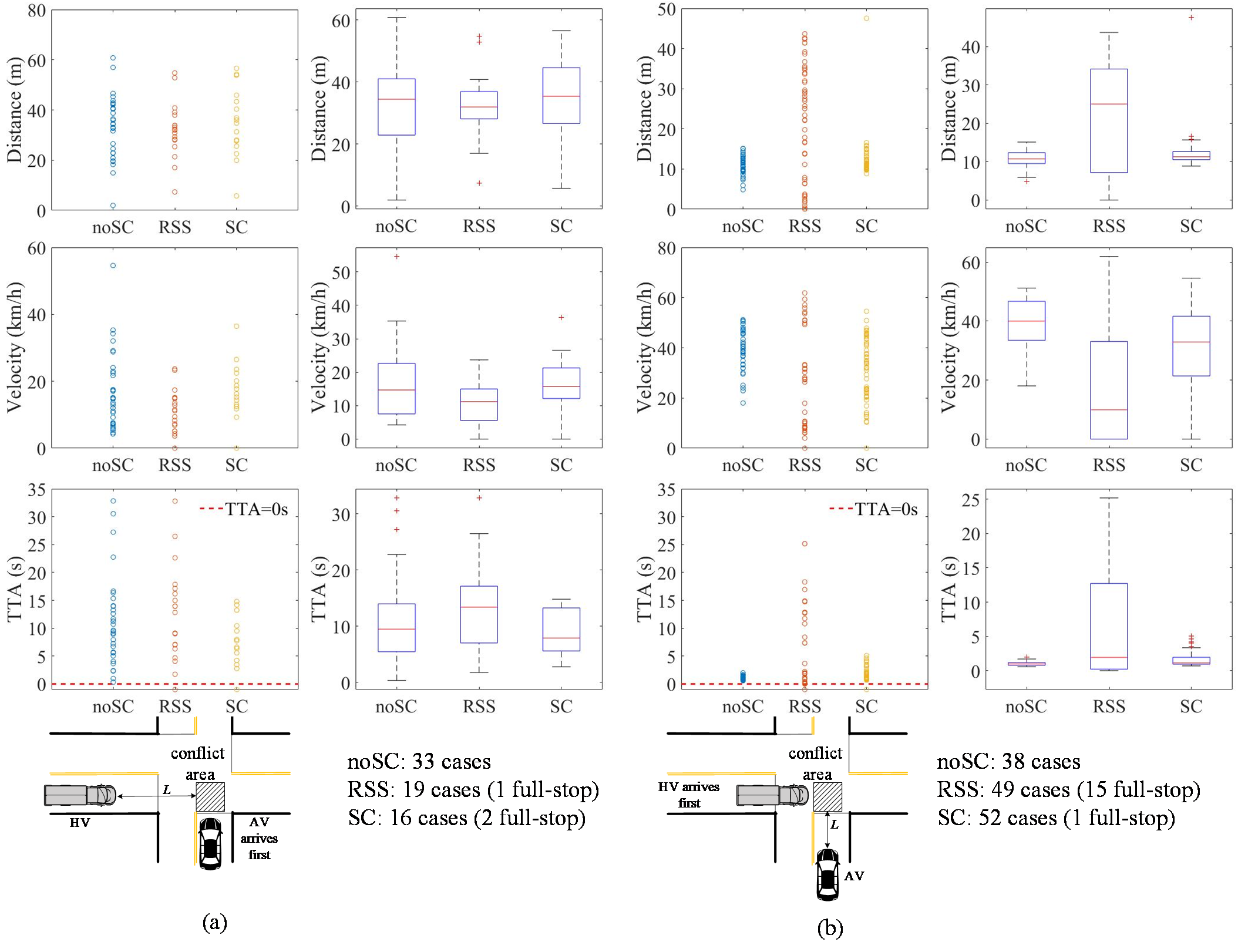}
  \caption{Statistics of the lagging vehicle's states at the end of interactions. a) Statistics of HV, when AV arrives first at conflict area. b) Statistics of AV, when HV arrives first at conflict area.}\label{fig10}

\end{figure}

Statistics of all interaction cases are summarized in Table \ref{tab3}. In the low speed scenarios, there are 6 danger cases with the noSC algorithm, and no danger case with RSS or SC algorithms. In medium and high speed scenarios, the numbers of danger cases with noSC, RSS and SC algorithms are 9, 11 and 6, respectively. The RSS algorithm seems conservative by showing the most courtesy behaviors, i.e. 16 full-stop cases. However, it still causes 11 danger cases. Therefore, although RSS is not responsible for any collisions, i.e. the interacting HVs bear the responsibility, it is not the safest algorithm for the studied intersection driving scenarios. By contrast, the SC algorithm can achieve the best safety performances in AV-HV interactions, with no danger case with initial speeds below $40km/h$ and 6 danger cases with initial speeds between 40 and 70$km/h$. 

\begin{table}[h!]
    \caption{Statistics of interaction cases (Danger: AEB activated, Full Stop: Yield)}
    \label{tab3}
    \begin{tabular}{@{}cccccc@{}}
    \hline
    \multicolumn{2}{c}{\# of cases}   & Low      & Low-Mid     & Mid        & High      \\
    \hline
    \multirow{3}{*}{noSC} & Total     & 22       & 15          & 26         & 8         \\
                          \cline{2-6}
                          & Danger       & \multicolumn{2}{c}{6}  & \multicolumn{2}{c}{9}  \\
                          \cline{2-6}
                          & Full Stop & \multicolumn{4}{c}{0}                           \\
                          \midrule
    \multirow{3}{*}{RSS}  & Total     & 20       & 12          & 27         & 9         \\
                          \cline{2-6}
                          & Danger       & \multicolumn{2}{c}{0}  & \multicolumn{2}{c}{11} \\
                          \cline{2-6}
                          & Full Stop & \multicolumn{4}{c}{16}                          \\
                          \midrule
    \multirow{3}{*}{SC}   & Total     & 17       & 15         & 25         & 11 \\
                          \cline{2-6}
                          & Danger       & \multicolumn{2}{c}{0}  & \multicolumn{2}{c}{6}  \\
                          \cline{2-6}
                          & Full Stop & \multicolumn{4}{c}{3}                           \\
    \hline
    \end{tabular}
    \end{table}

In 207 effective cases, the distance $L$, speed $v$ and TTA of the lagging vehicle at the end of the interactions are summarized in Figure \ref{fig10}. The cases when the AV or the HV arrives at the conflict area first are given in Figure \ref{fig10}a and Figure \ref{fig10}b, respectively. As shown in Figure \ref{fig10}a, when interacting with the RSS-based AV, the HV has the lowest ending velocity and its average TTA is larger than 10s, meaning that HV is the most conservative with lowest traffic efficiency. When interacting with the noSC-based AV, the efficiency of HV is improved, but there are some extremely conservative or radical cases, that is, its TTA is either too large or too small. When interacting with the SC-based AV, the average TTA of HV is the lowest, indicating the best traffic efficiency. Also, the SC-based AV facilitates the interacting HV to have its lower bound of TTA larger than that with the other two algorithms, showing its best safety performance of HV.

For the cases when HV arrives at the conflict area first, shown in Figure \ref{fig10}b, it can be found that RSS-based AV is the most conservative, having the lowest traffic efficiency among all algorithms. The widely-distributed TTA values indicate that RSS performs not stably or consistently in interacting with human drivers. Part of the reason is that RSS decides with a strict sense of right of the way (RoW), which may not always be precisely followed by human drivers. In highly-dynamic interactions, human drivers are not sensitive enough of their RoW. Such problem is getting worse if a human driver has visual limitations in sensing other interacting vehicles approaching the intersection, as the truck drivers in our experiments. This unclear sense of RoW may lead to the ineffective communication between HV and AV, causing the RSS-based AV to switch frequently between 'To Go' and 'Not to Go' decisions. Based on the results of TTA and ending velocity distribution of AVs in Figure \ref{fig10}b, the SC algorithm can provide the AV with the best tradeoff between safety and efficiency.

\subsection{Interaction case studies}
\label{section_casestudies}

To explain the benefits of social compatibility, we select three interaction cases with similar initial speeds, i.e. $52.0km/h$(noSC), $54.3km/h$(RSS) and $50.1km/h$(SC). For the HV in all three cases, the driver’s throttle input fluctuates between $30\%\sim45\%$, the vehicle acceleration fluctuates between $0.1\sim0.15m/s^2$, and the speed increment is between $3\sim4km/h$. 

\begin{figure}[h!]
  \includegraphics[width=12cm]{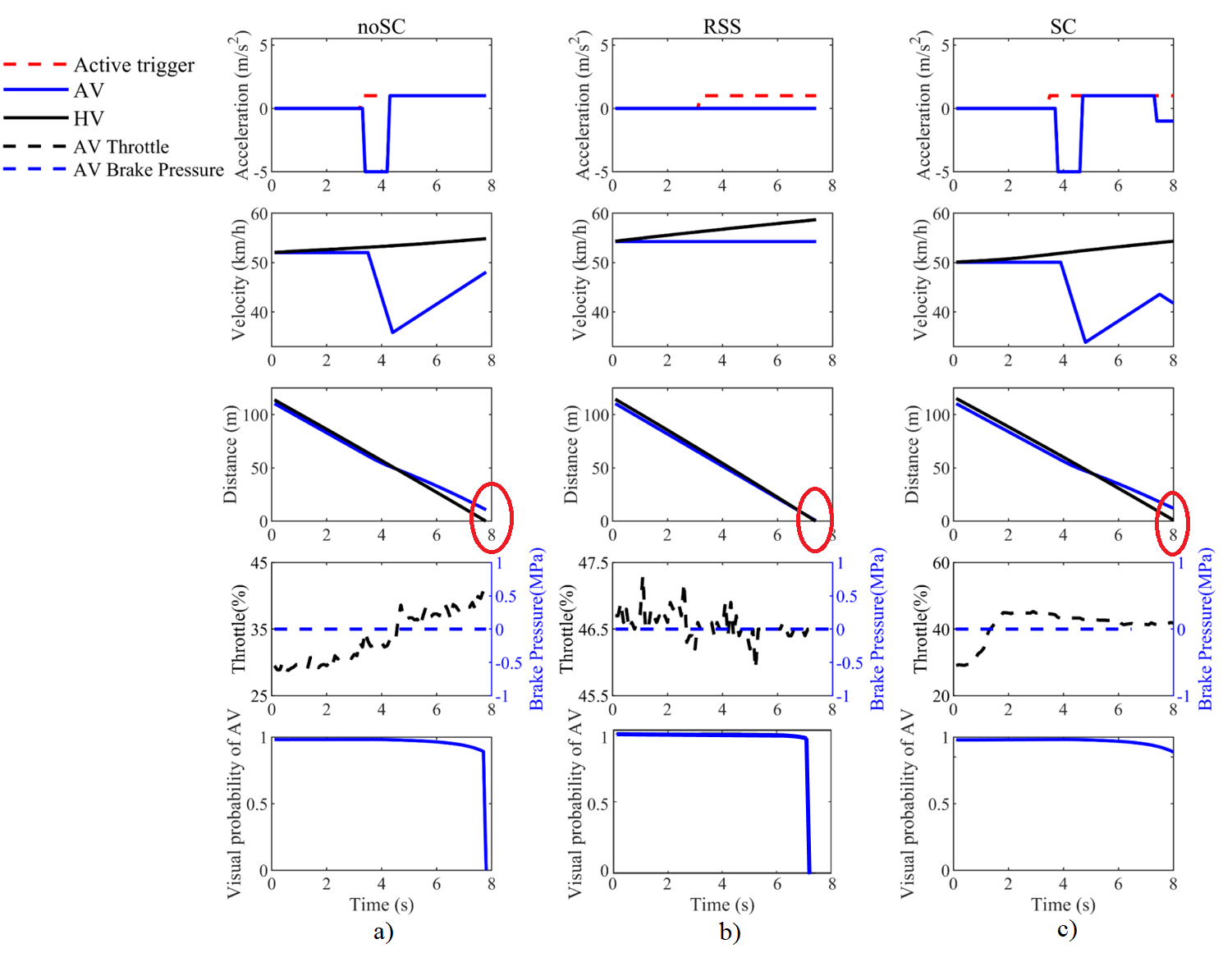}
  \caption{Comparisons of AV-HV interaction processes with three algorithms. a) noSC algorithm. b) RSS algorithm. c) the proposed SC algorithm.}\label{fig11}
    
\end{figure}

Figure \ref{fig11} presents the two vehicles’ states, the AV inputs, as well as the AV visibility probability estimated using \ref{eq2}. Since the noSC algorithm cannot consider the influence of HV visual limitations, the AV enters the blind zone of HV driver at the end of the interaction, with its visibility probability dropping to 0. The resulting TTA is $0.79s$, which is less than the specified threshold of 0.83 and triggers the AEB braking. 

As for the RSS case, the AV maintains the no-braking strategy according to the principle of right of way priority, but the HV does not slow down and yield according to the rules of the RSS, which finally leads to an almost inevitable collision (TTA=$0.02s$). When the RSS-based AV is close to the intersection, it enters the blind zone of the HV driver, and its visibility probability drops to 0. Therefore, if HV follows the rule of right of way priority, the RSS algorithm can achieve a safe interaction, otherwise a collision accident may happen. In the latter case, the RSS algorithm still do not lead to any accidents of AV's blame, but the accident still happens, showing that RSS is pursuing an egoism strategy and needs improvements for interaction-intensive driving. 

By contrast, since the SC algorithm can directly consider the HV driver’s visual limitations, the SC-based AV decelerates at $t=7.5s$ to keep away from the blind zone of HV driver. At the end of the interaction, its visual probability is 0.85, which is still a high probability of AV being observed by the HV driver. {At the end of the interaction}, TTA is $0.95s$, meaning a safe and efficient interaction with HV.

Together with the statistical results in Section \ref{section_statistics}, it is validated that by considering the social compatibility from the perspective of HV driver's visual limitation, the proposed decision algorithm can achieve both safety and efficiency.

\subsection{Subjective evaluation}

\begin{figure}[h!]
  \includegraphics[width=12cm]{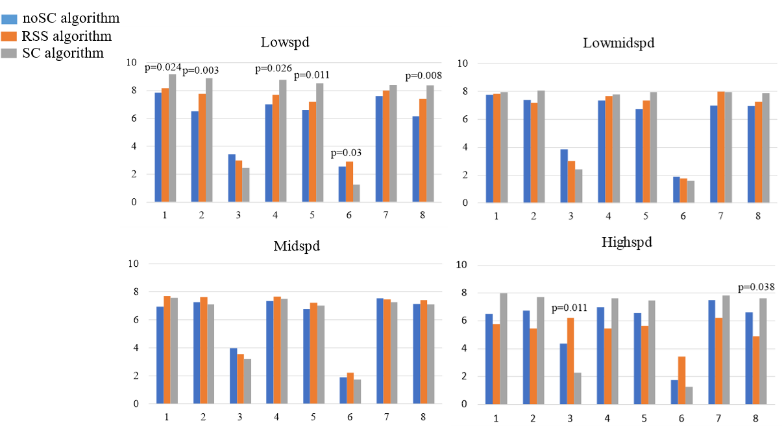}
  \caption{HV drivers' subjective evaluation on AV-HV interactions at different speeds. For each of 8 items of evaluation, the mean scores and significance values are given.}
  \label{fig12}
    
\end{figure}

The HV drivers' evaluations on AV are obtained via questionnaires in Table B.1. Figure \ref{fig12} presents the mean and significance values of subjective evaluation on interactions, with all significance levels ($p<0.05$) indicated for corresponding question items. {It shows that,} in all speed scenarios, the SC algorithm has better evaluation scores than the noSC algorithm in all items but item 2 \emph{'comfort'} and item 7 \emph{'calmness'}. In \emph{High} speed scenarios, the SC algorithm performs better in all items than the noSC and RSS algorithms, with two items with significant improvements. By contrast, in \emph{Low} speed scenarios, the SC algorithm shows the most significant improvements in 6 evaluation items. One possible reason is that if given a specified decision step size, there are more frequent interactions in a lower-speed interaction case. This makes the subjects easier to tell the differences among the three decision algorithms, so the advantages of the SC algorithm are more obvious.

\begin{figure}[h!]
  \includegraphics[width=12cm]{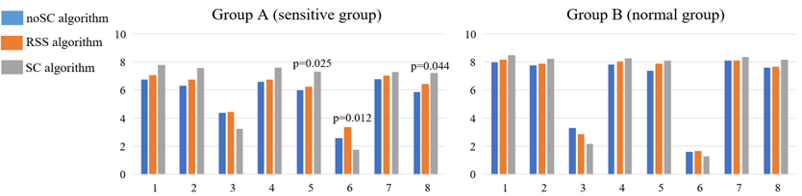}
  \caption{The sensitive and normal drivers' subjective evaluation on AV-HV interactions. For each of 8 items of evaluation, the mean scores and significance values are given.}
    
  \label{fig13}
\end{figure}

When examining those subjects who gave significantly unsatisfactory evaluations than average, we find that they have one or more of the following characteristics in driver self-ability assessment, i.e. poor driving ability, aggressive driving style, being prone to anger, or careless driving. A total of 12 subjects with the above characteristics are named Group A (sensitive), and the rest are classified into Group B (normal). The mean values and significance results of Groups A and B are shown in Figure \ref{fig13}. With the SC algorithm, the evaluation items 5 \emph{'relaxed'}, 6 \emph{'confused'} and 8 \emph{'happy'} are significantly improved for the sensitive subjects. By contrast, such improvements are not statistically significant for the normal subjects. It may be because Group A subjects are more sensitive to the process of dynamic interaction, and their mood fluctuations are more susceptible to the driving behaviors of interacting vehicle.

\subsection{HV driver EEG}
The EEG data at Fz and Cz are tagged with three stages. The \emph{Baseline} data correspond to the stage before the AV decision algorithm is activated ON, which are considered as the EEG data before the interaction. The \emph{Interaction} data correspond to the stage of interaction, i.e. when the HV drives from 120m away from the conflict area to the end of interaction. The \emph{After} data correspond to the stage of 6s after the interaction. For the EEG signal features, the mean power values of Alpha ($8\sim13$Hz), Beta ($13\sim30$Hz) and Theta ($4\sim8$Hz) waves are extracted to judge the subjects’ emotion fluctuation. {For each of the total 24 subjects}, the EEG results during interactions for one speed limit are taken as a data group. Finally, 61 effective data groups are obtained by eliminating the failed group, including those with over speed driving or lost EEG signals.

Cao \cite{ss18} pointed out that the power of Alpha and Theta waves increases when user feels more pleasure, and the power of Beta wave rises with the enhancing of positive emotions. Here, for the 61 effective data groups, with the power analysis of Alpha, Theta, and Beta waves, it is found that in 44 data groups, i.e. 72\%, the EEG evidences can confirm the driver emotion changes represented by the corresponding subjective evaluation (items 2, 4, 5 and 8). In the rest of 17 data groups, the EEG results are not consistent with the subjective evaluation.

\begin{figure}[h!]
  \includegraphics[width=12cm]{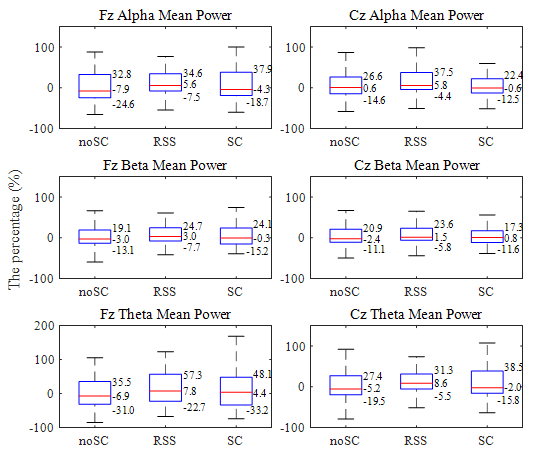}
  \caption{EEG mean power variation statistics of HV driver during AV-HV interactions.}
    
  \label{fig14}
\end{figure}

For the 61 effective groups of EEG data, the variation percentage of EEG mean power in each interaction is defined as $GR=(P_{base}-P_{int})/P_{base}*100\%$, where $P_{base}$ represents the mean power in the \emph{Baseline} stage, $P_{int}$ represents the mean power in the \emph{Interaction} stage. The statistical results are shown in the Figure \ref{fig14}. It is found that when subjects interact with the AVs with the RSS and SC algorithms, the mean power values of all EEG features are higher than those with the noSC algorithm. This confirms the subjective evaluation results that when interacting with the noSC algorithm, the satisfaction level of the subjects is lowest. By contrast, by considering social compatibility, the SC algorithm can provide an equivalent level of satisfaction as the conservative RSS algorithm.

The findings of human-in-the-loop experiments of AV-HV interactions can be summarized as follow.
\begin{enumerate}
\item Compared to the other benchmarks, the proposed SC algorithm can better balance safety and traffic efficiency, and achieve smoother interactions between AV and HV. 
\item From the microscopic case studies, the consideration of human visual limitations and social compatibility can help avoid less effective interactions due to blind zones, which can better improve safety. 
\item When the AV with our SC algorithm interacts with human drivers, in addition to objective performances of safety and efficiency, it improves its own predictability and makes the HV drivers feel safer and clearer about the inter-vehicle interactions. This is a significant improvement over the commonly-used algorithms in current AVs. This further confirms that to be a real safe and trust-worthy traffic participant, AV should not only make decisions primly according to safety rules and the right of way, but also behave empathetically by considering other human drivers' limits of driving capabilities.
\end{enumerate}

\section{Conclusion}
\label{Conclusion}

{The aim of this study is to propose an unsignalized intersection decision algorithm that can achieve safety, efficiency and social compatibility during dynamic interactions with human-driven vehicles.} 
\begin{enumerate}
\item A probabilistic model of the interacting driver’s visual limitations is constructed, which can estimate the probability of AV being observed by the human driver during the interaction process. 

\item Based on this visibility model, social compatibility is further realized using a game-theoretic framework. 

\item Human-in-the-loop experiments are carried out for the validation of the proposed algorithm. Results show that in addition to the well-balanced safety and time efficiency, the proposed AV decision algorithm can significantly improve social compatibility and make AV decision more in line with the expectation of human drivers.
\end{enumerate}

This study is one step further towards more advanced and human-like decision algorithms for automated vehicles. However, here we focus on realizing social compatibility from the perspective of other drivers' visual perception, while in future work the AV visibility model can be improved by considering the interaction uncertainties. Additionally, the main idea of incorporating social compatibility in AV decisions may be further applied in other interacting driving scenarios.

\appendix
\section*{Appendix I}\label{Appendix}
\subsection*{A. Detailed utility functions of AV and HV}\label{AppendixA}

\subsubsection*{AV utility}

\begin{figure}[h!]
  \includegraphics[width=8cm]{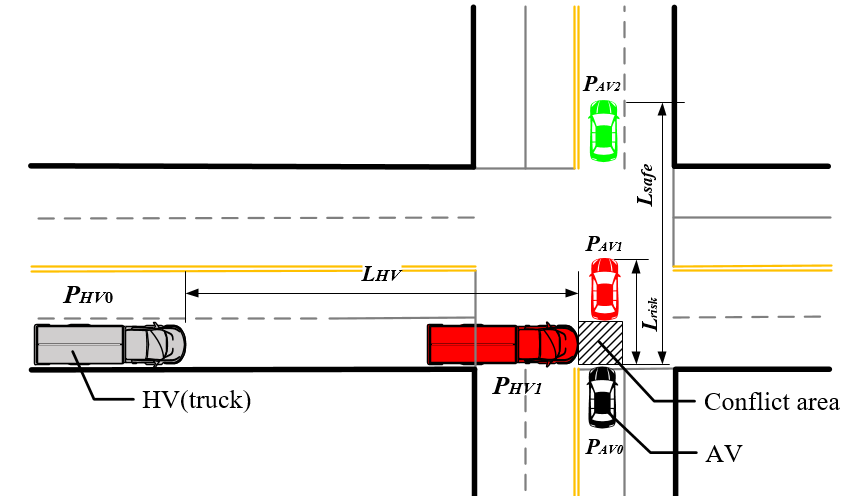}
  \caption{Intersection driving description.}\label{fig15}

\end{figure}

For a given interaction process, $\Delta t$ is defined as the time difference between the AV and HV’s arriving time at the conflict area. By normalizing the time difference $\Delta t$, we can describe the safety utility $u_s$ of a two-vehicle interaction process as follows, which is a value between -1 and 1. The safety utility of AV is $u_{s,AV}=u_s$, while the safety utility of HV depends on how well the AV can be observed by the HV driver.

\begin{equation}
\label{eq7}
f_i(\theta)=\left\{
      \begin{array}{cl}
        \Delta t / \Delta t_{rsk} - 1, & \Delta t \in [0,\Delta t_{rsk}]  \\
        (\Delta t - \Delta t_{rsk}) / (\Delta t_{saf} - \Delta t_{rsk}), & \Delta t \in (\Delta t_{rsk},\Delta t_{saf}) \\
        1, & \Delta t \in [\Delta t_{saf}, +\infty) 
      \end{array}
\right.
\end{equation}
where the parameters $\Delta t_{rsk}$ and $\Delta t_{saf}$ are the risky and safe thresholds of time difference $\Delta t$, respectively. As shown in Figure \ref{fig15}, the overlapped path of interactive vehicles is defined as the conflict area. At time $t_0$, if AV arrives at the conflict area first, given HV’s location $P_{HV0}$, velocity $v_{HV}$, the distance to conflict area $L_{HV}$, the thresholds {$\Delta t_{rsk}$,$\Delta t_{saf}$} are determined as follow. If when HV arrives at conflict area (location $P_{HV1}$), AV has just left conflict area meanwhile (location $P_{AV1}$), this time difference is defined as $\Delta t_{rsk}$. On the other hand, if AV has left the intersection (location $P_{AV2}$), this time difference is defined as $\Delta t_{saf}$, as shown in Eq. \ref{eq10}. Similarly, if HV will arrive at the conflict area at time $t_0$, we can also calculate the corresponding safety utility.

\begin{equation}
\label{eq8}
\left\{
      \begin{array}{ll}
        \Delta t & = L_{HV} / v_{HV}  \\
        \Delta t_{rsk} & = L_{risk} / v_{HV} \\
        \Delta t_{saf} & = L_{safe} / v_{HV} \\ 
      \end{array}
\right.
\end{equation}

Assuming the AV distance from the conflict area at time $t$ is $L_{AV}$, and the velocity is $v_{AV}$, then $t_{AV}=L_{AV}/v_{AV}$. If setting the maximum allowable velocity is $v_{max}$, an efficiency time is defined as $t_{eff,AV}=L_{AV}/v_{max}$. Then the traffic efficiency utility of AV, $u_{t,AV}$ is

\begin{equation}
\label{eq9}
u_{t,AV}=\left\{
      \begin{array}{cl}
        1-(t_{AV}-t_{eff,AV})/t_{eff,AV}, & t_{AV} <= t_{eff, AV} \\
        1, & t_{AV} > t_{eff, AV} 
      \end{array}
\right.
\end{equation}

The social fitness utility of AV $u_{sf,AV}$ represents how much the AV decision fits to the HV decision, which is modelled in Eq. \ref{eq10} with the AV visibility probability $F(\theta)$ and the tacitness degree $f_{tacit}(i,j)$. If $F(\theta)$ is small, the HV driver can hardly notice AV, so there is no cooperative driving behavior between them. The degree of tacit cooperation is explained in Table \ref{tab4}, in which $(i,j)$ stand for AV and HV, respectively. When HV adopts the Yield strategy, if AV yields as well, the tacitness degree is $f_{tacit}=0$, if AV does not yield, we set $f_{tacit}=1$.

\begin{equation}
\label{eq10}
u_{sf,AV}=F(\theta)f_{tacit}.
\end{equation}
  
\begin{table}[h!]
      \label{tab4}
 \caption{The tacitness degree $f_{tacit}$ of AV under different conditions}
    
    \begin{tabular}{@{}llll@{}} 
      \hline
      \multicolumn{2}{c}{\multirow{2}{*}{$f_{tacit}$}} & \multicolumn{2}{c}{HV} \\ \cline{3-4}
      \multicolumn{2}{c}{}                             & Yield    & Not Yield   \\
      \hline
      \multirow{2}{*}{AV}          & Yield             & 0        & 1           \\
                                   & Not Yield         & 1        & 0           \\
      \hline
      \end{tabular}
      \end{table}

\subsubsection*{HV utility}

The safety utility of HV, $u_{s,HV}$ is designed as

\begin{equation}
\label{eq11}
u_{s,HV}=\left\{
      \begin{array}{cl}
        (u_s)^{F(\theta)}, & u_s >= 0 \\
        (-u_s)^{F(\theta)}, & u_s < 0 
      \end{array}
\right.
\end{equation}
where the AV visibility probability $F(\theta)$ is introduced to correct the safety utility $u_s$. For example, when AV is in the blind zones or is almost invisible from the perspective of HV driver, it is assumed that there is no vehicle interacting with HV, and the maximum safety utility is achieved, $u_{s,HV}=1$.

Similar to Eq. \ref{eq11}, the traffic efficiency utility of HV is as follows.

\begin{equation}
\label{eq12}
u_{t,HV}=\left\{
      \begin{array}{cl}
        1-(t_{HV}-t_{eff,HV})/t_{eff,HV}, & t_{HV} <= t_{eff,HV} \\
        1, & t_{HV} > t_{eff,HV}
      \end{array}
\right.
\end{equation}

The reciprocal utility from the HV’s altruistic behavior is quantified with the traffic efficiency of AV and the AV visibility probability $F(\theta)$, i.e.

\begin{equation}
  \label{eq13}
  u_{altr,HV}(\theta)=F(\theta)u_{t,AV} 
\end{equation}

\subsection*{B. Questionnaire for the HV drivers' evaluation on the AV-HV interaction}\label{AppendixB}

Instruction for subject drivers:
"Please score the items in Table \ref{tab5} based on your feelings about your last interaction with the other vehicle. "

\begin{table}[h!]
\caption{Questionnaire for evaluation on the AV-HV interaction}\label{tab5}
        \begin{tabular}{@{}ll@{}}
        \hline
        Evaluation item                                       & Score                      \\
        \hline
        1. I feel safe in interation                          & \multirow{8}{*}{0$\sim$10} \\
        2. I feel comfortable in interation                   &                            \\
        3. I worry about collision with the other vehicle     &                            \\
        4. I feel satisfied with the interaction              &                            \\
        5. I feel relaxed in interaction                      &                            \\
        6. I am confused by the behavior of the other vehicle &                            \\
        7. I feel calm in interaction                         &                            \\
        8. I feel happy in interaction                        &                            \\
        \hline
        \end{tabular}
        \end{table}

\section*{Appendix II}


\begin{backmatter}


\section*{Funding}
This work was supported by Department of Science and Technology of Zhejiang (No. 2022C01241 and No. 2018C01058).


\section*{Availability of data and materials}
A video abstract is provided, which includes a brief introduction of the proposed approach and a video of experiments. Please visit http://b23.tv/xAOo2ib.

\section*{Ethics approval and consent to participate}
Ethics approval was obtained for this study, and all human subjects gave their informed consent.

\section*{Competing interests}
The authors declare that they have no competing interests.


\section*{Authors' contributions}
DL contributed in conceptualization, funding acquisition, algorithm, writing and video abstract. AL and HP contributed in algorithm, visualization, writing - review \& editing, and video abstract. WC was a major contributor in methodology, algorithm, data curation and writing the original draft.

\section*{Authors' information} 
DL received the B.S. degree in Vehicle Engineering from the Jilin University of Technology, Changchun, China, in 2003, and the Ph.D. degree in Vehicle Engineering from the Shanghai Jiao Tong University, Shanghai, China, in 2008.
He joined Zhejiang University in June 2008, first as a Post-Doc, then served as Assistant Professor, and now is Associate Professor. In 2011, he was a Visiting Scholar with the University of Missouri-Columbia, and from 2014 to 2016, he was a Visiting Scholar with the University of Michigan, Ann Arbor, Michigan.  He currently directs the Research Group of Human-Mobility-Automation (https://person.zju.edu.cn/daofei), and his research interests include vehicle dynamics and control, automated driving and complete human-vehicle system dynamics. AL, HP and WC are graduate students in the group.


\bibliographystyle{bmc-mathphys} 

 
\bibliography{CJME}     


\begin{thebibliography}{39}
\ifx \bisbn   \undefined \def \bisbn  #1{ISBN #1}\fi
\ifx \binits  \undefined \def \binits#1{#1}\fi
\ifx \bauthor  \undefined \def \bauthor#1{#1}\fi
\ifx \batitle  \undefined \def \batitle#1{#1}\fi
\ifx \bjtitle  \undefined \def \bjtitle#1{#1}\fi
\ifx \bvolume  \undefined \def \bvolume#1{\textbf{#1}}\fi
\ifx \byear  \undefined \def \byear#1{#1}\fi
\ifx \bissue  \undefined \def \bissue#1{#1}\fi
\ifx \bfpage  \undefined \def \bfpage#1{#1}\fi
\ifx \blpage  \undefined \def \blpage #1{#1}\fi
\ifx \burl  \undefined \def \burl#1{\textsf{#1}}\fi
\ifx \doiurl  \undefined \def \doiurl#1{\textsf{#1}}\fi
\ifx \betal  \undefined \def \betal{\textit{et al.}}\fi
\ifx \binstitute  \undefined \def \binstitute#1{#1}\fi
\ifx \binstitutionaled  \undefined \def \binstitutionaled#1{#1}\fi
\ifx \bctitle  \undefined \def \bctitle#1{#1}\fi
\ifx \beditor  \undefined \def \beditor#1{#1}\fi
\ifx \bpublisher  \undefined \def \bpublisher#1{#1}\fi
\ifx \bbtitle  \undefined \def \bbtitle#1{#1}\fi
\ifx \bedition  \undefined \def \bedition#1{#1}\fi
\ifx \bseriesno  \undefined \def \bseriesno#1{#1}\fi
\ifx \blocation  \undefined \def \blocation#1{#1}\fi
\ifx \bsertitle  \undefined \def \bsertitle#1{#1}\fi
\ifx \bsnm \undefined \def \bsnm#1{#1}\fi
\ifx \bsuffix \undefined \def \bsuffix#1{#1}\fi
\ifx \bparticle \undefined \def \bparticle#1{#1}\fi
\ifx \barticle \undefined \def \barticle#1{#1}\fi
\ifx \bconfdate \undefined \def \bconfdate #1{#1}\fi
\ifx \botherref \undefined \def \botherref #1{#1}\fi
\ifx \url \undefined \def \url#1{\textsf{#1}}\fi
\ifx \bchapter \undefined \def \bchapter#1{#1}\fi
\ifx \bbook \undefined \def \bbook#1{#1}\fi
\ifx \bcomment \undefined \def \bcomment#1{#1}\fi
\ifx \oauthor \undefined \def \oauthor#1{#1}\fi
\ifx \citeauthoryear \undefined \def \citeauthoryear#1{#1}\fi
\ifx \endbibitem  \undefined \def \endbibitem {}\fi
\ifx \bconflocation  \undefined \def \bconflocation#1{#1}\fi
\ifx \arxivurl  \undefined \def \arxivurl#1{\textsf{#1}}\fi
\csname PreBibitemsHook\endcsname

\bibitem{kyriakidis2019}
\begin{barticle}
\bauthor{\bsnm{Kyriakidis}, \binits{M.}},
\bauthor{\bparticle{de} \bsnm{Winter}, \binits{J.C.}},
\bauthor{\bsnm{Stanton}, \binits{N.}},
\bauthor{\bsnm{Bellet}, \binits{T.}},
\bauthor{\bparticle{van} \bsnm{Arem}, \binits{B.}},
\bauthor{\bsnm{Brookhuis}, \binits{K.}},
\bauthor{\bsnm{Martens}, \binits{M.H.}},
\bauthor{\bsnm{Bengler}, \binits{K.}},
\bauthor{\bsnm{Andersson}, \binits{J.}},
\bauthor{\bsnm{Merat}, \binits{N.}}, \betal:
\batitle{A human factors perspective on automated driving}.
\bjtitle{Theoretical Issues in Ergonomics Science}
\bvolume{20}(\bissue{3}),
\bfpage{223}--\blpage{249}
(\byear{2019})
\end{barticle}
\endbibitem

\bibitem{manchon2021}
\begin{barticle}
\bauthor{\bsnm{Manchon}, \binits{J.}},
\bauthor{\bsnm{Bueno}, \binits{M.}},
\bauthor{\bsnm{Navarro}, \binits{J.}}:
\batitle{From manual to automated driving: How does trust evolve?}
\bjtitle{Theoretical Issues in Ergonomics Science}
\bvolume{22}(\bissue{5}),
\bfpage{528}--\blpage{554}
(\byear{2021})
\end{barticle}
\endbibitem

\bibitem{Liusa2021}
\begin{botherref}
\oauthor{\bsnm{Liu}, \binits{S.}}:
Optimization of intelligent driving decision algorithm for trust enhancement.
Master {Thesis},
Zhejiang University
(2021)
\end{botherref}
\endbibitem

\bibitem{xu_when_2021}
\begin{barticle}
\bauthor{\bsnm{Xu}, \binits{Z.}},
\bauthor{\bsnm{Jiang}, \binits{Z.}},
\bauthor{\bsnm{Wang}, \binits{G.}},
\bauthor{\bsnm{Wang}, \binits{R.}},
\bauthor{\bsnm{Li}, \binits{T.}},
\bauthor{\bsnm{Liu}, \binits{J.}},
\bauthor{\bsnm{Zhang}, \binits{Y.}},
\bauthor{\bsnm{Liu}, \binits{P.}}:
\batitle{When the automated driving system fails: {Dynamics} of public
  responses to automated vehicles}.
\bjtitle{Transportation Research Part C: Emerging Technologies}
\bvolume{129},
\bfpage{103271}
(\byear{2021}).
doi:\doiurl{10.1016/j.trc.2021.103271}
\end{barticle}
\endbibitem

\bibitem{yu_measurement_2021}
\begin{barticle}
\bauthor{\bsnm{Yu}, \binits{B.}},
\bauthor{\bsnm{Bao}, \binits{S.}},
\bauthor{\bsnm{Zhang}, \binits{Y.}},
\bauthor{\bsnm{Sullivan}, \binits{J.}},
\bauthor{\bsnm{Flannagan}, \binits{M.}}:
\batitle{Measurement and prediction of driver trust in automated vehicle
  technologies: {An} application of hand position transition probability
  matrix}.
\bjtitle{Transportation Research Part C: Emerging Technologies}
\bvolume{124},
\bfpage{102957}
(\byear{2021}).
doi:\doiurl{10.1016/j.trc.2020.102957}
\end{barticle}
\endbibitem

\bibitem{Noy2022}
\begin{barticle}
\bauthor{\bsnm{Noy}, \binits{I.Y.}},
\bauthor{\bsnm{Shinar}, \binits{D.}},
\bauthor{\bsnm{Horrey}, \binits{W.J.}}:
\batitle{Automated driving: Safety blind spots}.
\bjtitle{Safety Science}
\bvolume{102},
\bfpage{68}--\blpage{78}
(\byear{2018}).
doi:\doiurl{10.1016/j.ssci.2017.07.018}
\end{barticle}
\endbibitem

\bibitem{li2021planning}
\begin{barticle}
\bauthor{\bsnm{Li}, \binits{S.}},
\bauthor{\bsnm{Shu}, \binits{K.}},
\bauthor{\bsnm{Chen}, \binits{C.}},
\bauthor{\bsnm{Cao}, \binits{D.}}:
\batitle{Planning and decision-making for connected autonomous vehicles at road
  intersections: A review}.
\bjtitle{Chinese Journal of Mechanical Engineering}
\bvolume{34}(\bissue{1}),
\bfpage{1}--\blpage{18}
(\byear{2021})
\end{barticle}
\endbibitem

\bibitem{Di2021TRC}
\begin{barticle}
\bauthor{\bsnm{Di}, \binits{X.}},
\bauthor{\bsnm{Shi}, \binits{R.}}:
\batitle{A survey on autonomous vehicle control in the era of mixed-autonomy:
  From physics-based to ai-guided driving policy learning}.
\bjtitle{Transportation Research Part C: Emerging Technologies}
\bvolume{125},
\bfpage{103008}
(\byear{2021}).
doi:\doiurl{10.1016/j.trc.2021.103008}
\end{barticle}
\endbibitem

\bibitem{ss2}
\begin{botherref}
\oauthor{\bsnm{Schwall}, \binits{M.}},
\oauthor{\bsnm{Daniel}, \binits{T.}},
\oauthor{\bsnm{Victor}, \binits{T.}},
\oauthor{\bsnm{Favaro}, \binits{F.}},
\oauthor{\bsnm{Hohnhold}, \binits{H.}}:
Waymo public road safety performance data.
arXiv
(2020).
doi:\doiurl{10.48550/ARXIV.2011.00038}
\end{botherref}
\endbibitem

\bibitem{ss23}
\begin{barticle}
\bauthor{\bsnm{Schwarting}, \binits{W.}},
\bauthor{\bsnm{Alonso-Mora}, \binits{J.}},
\bauthor{\bsnm{Rus}, \binits{D.}}:
\batitle{Planning and decision-making for autonomous vehicles}.
\bjtitle{Annual Review of Control, Robotics, and Autonomous Systems}
\bvolume{1}(\bissue{1}),
\bfpage{187}--\blpage{210}
(\byear{2018}).
doi:\doiurl{10.1146/annurev-control-060117-105157}
\end{barticle}
\endbibitem

\bibitem{ss5}
\begin{bchapter}
\bauthor{\bsnm{Beaucorps}, \binits{P.}},
\bauthor{\bsnm{Streubel}, \binits{T.}},
\bauthor{\bsnm{Verroust-Blondet}, \binits{A.}},
\bauthor{\bsnm{Nashashibi}, \binits{F.}},
\bauthor{\bsnm{Bradai}, \binits{B.}},
\bauthor{\bsnm{Resende}, \binits{P.}}:
\bctitle{Decision-making for automated vehicles at intersections adapting
  human-like behavior}.
In: \bbtitle{2017 {IEEE} {Intelligent} {Vehicles} {Symposium} ({IV})},
pp. \bfpage{212}--\blpage{217}.
\bpublisher{IEEE},
\blocation{Los Angeles, CA, USA}
(\byear{2017}).
doi:\doiurl{10.1109/IVS.2017.7995722}
\end{bchapter}
\endbibitem

\bibitem{ss6}
\begin{bchapter}
\bauthor{\bsnm{Chen}, \binits{J.}},
\bauthor{\bsnm{Yuan}, \binits{B.}},
\bauthor{\bsnm{Tomizuka}, \binits{M.}}:
\bctitle{Deep imitation learning for autonomous driving in generic urban
  scenarios with enhanced safety}.
In: \bbtitle{2019 {IEEE}/{RSJ} {International} {Conference} on {Intelligent}
  {Robots} and {Systems} ({IROS})},
pp. \bfpage{2884}--\blpage{2890}.
\bpublisher{IEEE},
\blocation{Macau, China}
(\byear{2019}).
doi:\doiurl{10.1109/IROS40897.2019.8968225}
\end{bchapter}
\endbibitem

\bibitem{ss7}
\begin{bchapter}
\bauthor{\bsnm{Sezer}, \binits{V.}},
\bauthor{\bsnm{Bandyopadhyay}, \binits{T.}},
\bauthor{\bsnm{Rus}, \binits{D.}},
\bauthor{\bsnm{Frazzoli}, \binits{E.}},
\bauthor{\bsnm{Hsu}, \binits{D.}}:
\bctitle{Towards autonomous navigation of unsignalized intersections under
  uncertainty of human driver intent}.
In: \bbtitle{2015 {IEEE}/{RSJ} {International} {Conference} on {Intelligent}
  {Robots} and {Systems} ({IROS})},
pp. \bfpage{3578}--\blpage{3585}.
\bpublisher{IEEE},
\blocation{Hamburg, Germany}
(\byear{2015}).
doi:\doiurl{10.1109/IROS.2015.7353877}
\end{bchapter}
\endbibitem

\bibitem{ss9}
\begin{bchapter}
\bauthor{\bsnm{Menendez-Romero}, \binits{C.}},
\bauthor{\bsnm{Sezer}, \binits{M.}},
\bauthor{\bsnm{Winkler}, \binits{F.}},
\bauthor{\bsnm{Dornhege}, \binits{C.}},
\bauthor{\bsnm{Burgard}, \binits{W.}}:
\bctitle{Courtesy behavior for highly automated vehicles on highway
  interchanges}.
In: \bbtitle{2018 {IEEE} {Intelligent} {Vehicles} {Symposium} ({IV})},
pp. \bfpage{943}--\blpage{948}.
\bpublisher{IEEE},
\blocation{Changshu}
(\byear{2018}).
doi:\doiurl{10.1109/IVS.2018.8500407}
\end{bchapter}
\endbibitem

\bibitem{ss8}
\begin{bchapter}
\bauthor{\bsnm{Wang}, \binits{W.-J.}}:
\bctitle{Decision and behavior planning for a self-driving vehicle at
  unsignalized intersections}.
In: \bbtitle{2020 {International} {Automatic} {Control} {Conference} ({CACS})},
pp. \bfpage{1}--\blpage{6}.
\bpublisher{IEEE},
\blocation{Hsinchu, Taiwan}
(\byear{2020}).
doi:\doiurl{10.1109/CACS50047.2020.9289738}
\end{bchapter}
\endbibitem

\bibitem{ss26}
\begin{barticle}
\bauthor{\bsnm{Lefkopoulos}, \binits{V.}},
\bauthor{\bsnm{Menner}, \binits{M.}},
\bauthor{\bsnm{Domahidi}, \binits{A.}},
\bauthor{\bsnm{Zeilinger}, \binits{M.N.}}:
\batitle{Interaction-aware motion prediction for autonomous driving: A multiple
  model kalman filtering scheme}.
\bjtitle{IEEE Robotics and Automation Letters}
\bvolume{6}(\bissue{1}),
\bfpage{80}--\blpage{87}
(\byear{2021}).
doi:\doiurl{10.1109/LRA.2020.3032079}
\end{barticle}
\endbibitem

\bibitem{ss28}
\begin{bchapter}
\bauthor{\bsnm{Yoon}, \binits{Y.}},
\bauthor{\bsnm{Yi}, \binits{K.}}:
\bctitle{Design of longitudinal control for autonomous vehicles based on
  interactive intention inference of surrounding vehicle behavior using long
  short-term memory}.
In: \bbtitle{2021 {IEEE} {International} {Intelligent} {Transportation}
  {Systems} {Conference} ({ITSC})},
pp. \bfpage{196}--\blpage{203}.
\bpublisher{IEEE},
\blocation{Indianapolis, IN, USA}
(\byear{2021}).
doi:\doiurl{10.1109/ITSC48978.2021.9564986}
\end{bchapter}
\endbibitem

\bibitem{ss29}
\begin{barticle}
\bauthor{\bsnm{Wang}, \binits{L.}},
\bauthor{\bsnm{Wu}, \binits{T.}},
\bauthor{\bsnm{Fu}, \binits{H.}},
\bauthor{\bsnm{Xiao}, \binits{L.}},
\bauthor{\bsnm{Wang}, \binits{Z.}},
\bauthor{\bsnm{Dai}, \binits{B.}}:
\batitle{Multiple contextual cues integrated trajectory prediction for
  autonomous driving}.
\bjtitle{IEEE Robotics and Automation Letters}
\bvolume{6}(\bissue{4}),
\bfpage{6844}--\blpage{6851}
(\byear{2021}).
doi:\doiurl{10.1109/LRA.2021.3094564}
\end{barticle}
\endbibitem

\bibitem{ss30}
\begin{barticle}
\bauthor{\bsnm{Zhang}, \binits{T.}},
\bauthor{\bsnm{Song}, \binits{W.}},
\bauthor{\bsnm{Fu}, \binits{M.}},
\bauthor{\bsnm{Yang}, \binits{Y.}},
\bauthor{\bsnm{Wang}, \binits{M.}}:
\batitle{Vehicle motion prediction at intersections based on the turning
  intention and prior trajectories model}.
\bjtitle{IEEE/CAA Journal of Automatica Sinica}
\bvolume{8}(\bissue{10}),
\bfpage{1657}--\blpage{1666}
(\byear{2021}).
doi:\doiurl{10.1109/JAS.2021.1003952}
\end{barticle}
\endbibitem

\bibitem{Karle2022}
\begin{botherref}
\oauthor{\bsnm{Karle}, \binits{P.}},
\oauthor{\bsnm{Geisslinger}, \binits{M.}},
\oauthor{\bsnm{Betz}, \binits{J.}},
\oauthor{\bsnm{Lienkamp}, \binits{M.}}:
Scenario understanding and motion prediction for autonomous vehicles - review
  and comparison.
IEEE Transactions on Intelligent Transportation Systems,
1--21
(2022).
doi:\doiurl{10.1109/TITS.2022.3156011}
\end{botherref}
\endbibitem

\bibitem{linan2020game}
\begin{botherref}
\oauthor{\bsnm{Li}, \binits{N.}},
\oauthor{\bsnm{Yao}, \binits{Y.}},
\oauthor{\bsnm{Kolmanovsky}, \binits{I.}},
\oauthor{\bsnm{Atkins}, \binits{E.}},
\oauthor{\bsnm{Girard}, \binits{A.R.}}:
Game-theoretic modeling of multi-vehicle interactions at uncontrolled
  intersections.
IEEE Transactions on Intelligent Transportation Systems
(2020)
\end{botherref}
\endbibitem

\bibitem{jin2020game}
\begin{bchapter}
\bauthor{\bsnm{Jin}, \binits{X.}},
\bauthor{\bsnm{Li}, \binits{K.}},
\bauthor{\bsnm{Jia}, \binits{Q.-S.}},
\bauthor{\bsnm{Xia}, \binits{H.}},
\bauthor{\bsnm{Bai}, \binits{Y.}},
\bauthor{\bsnm{Ren}, \binits{D.}}:
\bctitle{A game-theoretic reinforcement learning approach for adaptive
  interaction at intersections}.
In: \bbtitle{2020 Chinese Automation Congress (CAC)},
pp. \bfpage{4451}--\blpage{4456}
(\byear{2020}).
\bcomment{IEEE}
\end{bchapter}
\endbibitem

\bibitem{cai2021game}
\begin{bchapter}
\bauthor{\bsnm{Cai}, \binits{J.}},
\bauthor{\bsnm{Hang}, \binits{P.}},
\bauthor{\bsnm{Lv}, \binits{C.}}:
\bctitle{Game theoretic modeling and decision making for connected vehicle
  interactions at urban intersections}.
In: \bbtitle{2021 6th IEEE International Conference on Advanced Robotics and
  Mechatronics (ICARM)},
pp. \bfpage{874}--\blpage{880}
(\byear{2021}).
\bcomment{IEEE}
\end{bchapter}
\endbibitem

\bibitem{chandra2022gameplan}
\begin{botherref}
\oauthor{\bsnm{Chandra}, \binits{R.}},
\oauthor{\bsnm{Manocha}, \binits{D.}}:
Gameplan: Game-theoretic multi-agent planning with human drivers at
  intersections, roundabouts, and merging.
IEEE Robotics and Automation Letters
(2022)
\end{botherref}
\endbibitem

\bibitem{Chenwentao2021}
\begin{botherref}
\oauthor{\bsnm{Chen}, \binits{W.}}:
Research on autonomous driving decision algorithm considering social
  compatibility.
Master {Thesis},
Zhejiang University
(2021)
\end{botherref}
\endbibitem

\bibitem{Wangletian2021}
\begin{barticle}
\bauthor{\bsnm{Wang}, \binits{L.}},
\bauthor{\bsnm{Sun}, \binits{L.}},
\bauthor{\bsnm{Tomizuka}, \binits{M.}},
\bauthor{\bsnm{Zhan}, \binits{W.}}:
\batitle{Socially-compatible behavior design of autonomous vehicles with
  verification on real human data}.
\bjtitle{IEEE Robotics and Automation Letters}
\bvolume{6}(\bissue{2}),
\bfpage{3421}--\blpage{3428}
(\byear{2021}).
doi:\doiurl{10.1109/LRA.2021.3061350}
\end{barticle}
\endbibitem

\bibitem{Lidaofei2022lgm}
\begin{barticle}
\bauthor{\bsnm{Li}, \binits{D.}},
\bauthor{\bsnm{Liu}, \binits{G.}},
\bauthor{\bsnm{Xiao}, \binits{B.}}:
\batitle{Human-like driving decision at unsignalized intersections based on
  game theory}.
\bjtitle{Proceedings of the Institution of Mechanical Engineers, Part D:
  Journal of Automobile Engineering}
(\byear{2022}).
doi:\doiurl{10.1177/09544070221075423}
\end{barticle}
\endbibitem

\bibitem{LiTTRA2022}
\begin{barticle}
\bauthor{\bsnm{Li}, \binits{D.}},
\bauthor{\bsnm{Hao}, \binits{P.}}:
\batitle{Two-lane two-way overtaking decision model with driving style
  awareness based on a game-theoretic framework}.
\bjtitle{Transportmetrica A: Transport Science}
(\byear{2022}).
doi:\doiurl{10.1080/23249935.2022.2076755}
\end{barticle}
\endbibitem

\bibitem{ss3}
\begin{botherref}
\oauthor{\bsnm{Ladegård}, \binits{G.}}:
Forming strategic alliances: The role of social compatibility.
Ph.{D}. {Thesis},
Norwegian School of Economics and Business Administration
(1997)
\end{botherref}
\endbibitem

\bibitem{ss4}
\begin{botherref}
\oauthor{\bsnm{{Traffic accident video}}}:
Personal space of traffic accident video
(2020).
\url{https://www.acfun.cn/u/4075269 (accessed\texttildelow  Feb. 08, 2021)}
\end{botherref}
\endbibitem

\bibitem{Larsen2004}
\begin{barticle}
\bauthor{\bsnm{Larsen}, \binits{L.}}:
\batitle{Methods of multidisciplinary in-depth analyses of road traffic
  accidents}.
\bjtitle{Journal of Hazardous Materials}
\bvolume{111}(\bissue{1}),
\bfpage{115}--\blpage{122}
(\byear{2004}).
doi:\doiurl{10.1016/j.jhazmat.2004.02.019}.
\bcomment{A Selection of Papers from the JRC/ESReDA Seminar on Safety
  Investigation Accidents, Petten, The Netherlands, 12-13 May, 2003}
\end{barticle}
\endbibitem

\bibitem{ss19}
\begin{bchapter}
\bauthor{\bsnm{Fang}, \binits{J.}},
\bauthor{\bsnm{Yan}, \binits{D.}},
\bauthor{\bsnm{Qiao}, \binits{J.}},
\bauthor{\bsnm{Xue}, \binits{J.}},
\bauthor{\bsnm{Wang}, \binits{H.}},
\bauthor{\bsnm{Li}, \binits{S.}}:
\bctitle{Dada-2000: Can driving accident be predicted by driver attention?
  analyzed by a benchmark}.
In: \bbtitle{2019 IEEE Intelligent Transportation Systems Conference (ITSC)},
\bconflocation{Auckland, NZ},
pp. \bfpage{4303}--\blpage{4309}
(\byear{2019}).
doi:\doiurl{10.1109/ITSC.2019.8917218}
\end{bchapter}
\endbibitem

\bibitem{ss12}
\begin{botherref}
\oauthor{\bsnm{Shalev-Shwartz}, \binits{S.}},
\oauthor{\bsnm{Shammah}, \binits{S.}},
\oauthor{\bsnm{Shashua}, \binits{A.}}:
On a formal model of safe and scalable self-driving cars.
arXiv
(2018).
doi:\doiurl{10.48550/ARXIV.1708.06374}
\end{botherref}
\endbibitem

\bibitem{ss13}
\begin{barticle}
\bauthor{\bsnm{Li}, \binits{L.}},
\bauthor{\bsnm{Zhu}, \binits{X.}},
\bauthor{\bsnm{Dong}, \binits{X.}},
\bauthor{\bsnm{Ma}, \binits{Z.}}:
\batitle{A research on the collision avoidance strategy for autonomous
  emergency braking system}.
\bjtitle{Automotive Engineering}
\bvolume{37}(\bissue{2}),
\bfpage{168}--\blpage{174}
(\byear{2015})
\end{barticle}
\endbibitem

\bibitem{ss14}
\begin{botherref}
\oauthor{\bsnm{Lin}, \binits{W.}}:
Emotion recognition and application based on physiological signals.
Ph.{D}. {Thesis},
Zhejiang University
(2019)
\end{botherref}
\endbibitem

\bibitem{ss15}
\begin{barticle}
\bauthor{\bsnm{Lajunen}, \binits{T.}},
\bauthor{\bsnm{Summala}, \binits{H.}}:
\batitle{Driving experience, personality, and skill and safety-motive
  dimensions in drivers’self-assessments}.
\bjtitle{Personality and Individual Differences}
\bvolume{19}(\bissue{3}),
\bfpage{307}--\blpage{318}
(\byear{1995})
\end{barticle}
\endbibitem

\bibitem{ss16}
\begin{botherref}
\oauthor{\bsnm{Liu}, \binits{J.}}:
Analysis on lane changing trajectory under different driving style and design
  on assistant lane changing system.
Master’s {Thesis},
Changsha University of Science \& Technology
(2015)
\end{botherref}
\endbibitem

\bibitem{ss17}
\begin{botherref}
\oauthor{\bsnm{Sun}, \binits{Y.}}:
Study on discretionary lane-changing behavior on urban streets.
Master’s {Thesis},
Dalian University of Technology
(2017)
\end{botherref}
\endbibitem

\bibitem{ss18}
\begin{botherref}
\oauthor{\bsnm{Cao}, \binits{K.}}:
The research of the {EEG} frequency power features in three basic emotions.
Master’s {Thesis},
Tianjin Medical University
(2019)
\end{botherref}
\endbibitem

\end{thebibliography}

\newcommand{\BMCxmlcomment}[1]{}

\BMCxmlcomment{

<refgrp>

<bibl id="B1">
  <title><p>A human factors perspective on automated driving</p></title>
  <aug>
    <au><snm>Kyriakidis</snm><fnm>M</fnm></au>
    <au><snm>Winter</snm><fnm>JC</fnm></au>
    <au><snm>Stanton</snm><fnm>N</fnm></au>
    <au><snm>Bellet</snm><fnm>T</fnm></au>
    <au><snm>Arem</snm><fnm>B</fnm></au>
    <au><snm>Brookhuis</snm><fnm>K</fnm></au>
    <au><snm>Martens</snm><fnm>MH</fnm></au>
    <au><snm>Bengler</snm><fnm>K</fnm></au>
    <au><snm>Andersson</snm><fnm>J</fnm></au>
    <au><snm>Merat</snm><fnm>N</fnm></au>
    <au><cnm>others</cnm></au>
  </aug>
  <source>Theoretical Issues in Ergonomics Science</source>
  <publisher>Taylor \& Francis</publisher>
  <pubdate>2019</pubdate>
  <volume>20</volume>
  <issue>3</issue>
  <fpage>223</fpage>
  <lpage>-249</lpage>
</bibl>

<bibl id="B2">
  <title><p>From manual to automated driving: How does trust
  evolve?</p></title>
  <aug>
    <au><snm>Manchon</snm><fnm>JB</fnm></au>
    <au><snm>Bueno</snm><fnm>M</fnm></au>
    <au><snm>Navarro</snm><fnm>J</fnm></au>
  </aug>
  <source>Theoretical Issues in Ergonomics Science</source>
  <publisher>Taylor \& Francis</publisher>
  <pubdate>2021</pubdate>
  <volume>22</volume>
  <issue>5</issue>
  <fpage>528</fpage>
  <lpage>-554</lpage>
</bibl>

<bibl id="B3">
  <title><p>Optimization of intelligent driving decision algorithm for trust
  enhancement</p></title>
  <aug>
    <au><snm>Liu</snm><fnm>S</fnm></au>
  </aug>
  <source>PhD thesis</source>
  <publisher>Zhejiang University</publisher>
  <pubdate>2021</pubdate>
</bibl>

<bibl id="B4">
  <title><p>When the automated driving system fails: {Dynamics} of public
  responses to automated vehicles</p></title>
  <aug>
    <au><snm>Xu</snm><fnm>Z</fnm></au>
    <au><snm>Jiang</snm><fnm>Z</fnm></au>
    <au><snm>Wang</snm><fnm>G</fnm></au>
    <au><snm>Wang</snm><fnm>R</fnm></au>
    <au><snm>Li</snm><fnm>T</fnm></au>
    <au><snm>Liu</snm><fnm>J</fnm></au>
    <au><snm>Zhang</snm><fnm>Y</fnm></au>
    <au><snm>Liu</snm><fnm>P</fnm></au>
  </aug>
  <source>Transportation Research Part C: Emerging Technologies</source>
  <pubdate>2021</pubdate>
  <volume>129</volume>
  <fpage>103271</fpage>
</bibl>

<bibl id="B5">
  <title><p>Measurement and prediction of driver trust in automated vehicle
  technologies: {An} application of hand position transition probability
  matrix</p></title>
  <aug>
    <au><snm>Yu</snm><fnm>B</fnm></au>
    <au><snm>Bao</snm><fnm>S</fnm></au>
    <au><snm>Zhang</snm><fnm>Y</fnm></au>
    <au><snm>Sullivan</snm><fnm>J</fnm></au>
    <au><snm>Flannagan</snm><fnm>M</fnm></au>
  </aug>
  <source>Transportation Research Part C: Emerging Technologies</source>
  <pubdate>2021</pubdate>
  <volume>124</volume>
  <fpage>102957</fpage>
</bibl>

<bibl id="B6">
  <title><p>Automated driving: Safety blind spots</p></title>
  <aug>
    <au><snm>Noy</snm><fnm>IY</fnm></au>
    <au><snm>Shinar</snm><fnm>D</fnm></au>
    <au><snm>Horrey</snm><fnm>WJ</fnm></au>
  </aug>
  <source>Safety Science</source>
  <pubdate>2018</pubdate>
  <volume>102</volume>
  <fpage>68</fpage>
  <lpage>-78</lpage>
</bibl>

<bibl id="B7">
  <title><p>Planning and Decision-making for Connected Autonomous Vehicles at
  Road Intersections: A Review</p></title>
  <aug>
    <au><snm>Li</snm><fnm>S</fnm></au>
    <au><snm>Shu</snm><fnm>K</fnm></au>
    <au><snm>Chen</snm><fnm>C</fnm></au>
    <au><snm>Cao</snm><fnm>D</fnm></au>
  </aug>
  <source>Chinese Journal of Mechanical Engineering</source>
  <publisher>SpringerOpen</publisher>
  <pubdate>2021</pubdate>
  <volume>34</volume>
  <issue>1</issue>
  <fpage>1</fpage>
  <lpage>-18</lpage>
</bibl>

<bibl id="B8">
  <title><p>A survey on autonomous vehicle control in the era of
  mixed-autonomy: From physics-based to AI-guided driving policy
  learning</p></title>
  <aug>
    <au><snm>Di</snm><fnm>X</fnm></au>
    <au><snm>Shi</snm><fnm>R</fnm></au>
  </aug>
  <source>Transportation Research Part C: Emerging Technologies</source>
  <pubdate>2021</pubdate>
  <volume>125</volume>
  <fpage>103008</fpage>
</bibl>

<bibl id="B9">
  <title><p>Waymo public road safety performance data</p></title>
  <aug>
    <au><snm>Schwall</snm><fnm>M</fnm></au>
    <au><snm>Daniel</snm><fnm>T</fnm></au>
    <au><snm>Victor</snm><fnm>T</fnm></au>
    <au><snm>Favaro</snm><fnm>F</fnm></au>
    <au><snm>Hohnhold</snm><fnm>H</fnm></au>
  </aug>
  <publisher>arXiv</publisher>
  <pubdate>2020</pubdate>
</bibl>

<bibl id="B10">
  <title><p>Planning and decision-making for autonomous vehicles</p></title>
  <aug>
    <au><snm>Schwarting</snm><fnm>W</fnm></au>
    <au><snm>Alonso Mora</snm><fnm>J</fnm></au>
    <au><snm>Rus</snm><fnm>D</fnm></au>
  </aug>
  <source>Annual Review of Control, Robotics, and Autonomous Systems</source>
  <pubdate>2018</pubdate>
  <volume>1</volume>
  <issue>1</issue>
  <fpage>187</fpage>
  <lpage>-210</lpage>
</bibl>

<bibl id="B11">
  <title><p>Decision-making for automated vehicles at intersections adapting
  human-like behavior</p></title>
  <aug>
    <au><snm>Beaucorps</snm><fnm>P</fnm></au>
    <au><snm>Streubel</snm><fnm>T</fnm></au>
    <au><snm>Verroust Blondet</snm><fnm>A</fnm></au>
    <au><snm>Nashashibi</snm><fnm>F</fnm></au>
    <au><snm>Bradai</snm><fnm>B</fnm></au>
    <au><snm>Resende</snm><fnm>P</fnm></au>
  </aug>
  <source>2017 {IEEE} {Intelligent} {Vehicles} {Symposium} ({IV})</source>
  <publisher>Los Angeles, CA, USA: IEEE</publisher>
  <pubdate>2017</pubdate>
  <fpage>212</fpage>
  <lpage>-217</lpage>
</bibl>

<bibl id="B12">
  <title><p>Deep imitation learning for autonomous driving in generic urban
  scenarios with enhanced safety</p></title>
  <aug>
    <au><snm>Chen</snm><fnm>J</fnm></au>
    <au><snm>Yuan</snm><fnm>B</fnm></au>
    <au><snm>Tomizuka</snm><fnm>M</fnm></au>
  </aug>
  <source>2019 {IEEE}/{RSJ} {International} {Conference} on {Intelligent}
  {Robots} and {Systems} ({IROS})</source>
  <publisher>Macau, China: IEEE</publisher>
  <pubdate>2019</pubdate>
  <fpage>2884</fpage>
  <lpage>-2890</lpage>
</bibl>

<bibl id="B13">
  <title><p>Towards autonomous navigation of unsignalized intersections under
  uncertainty of human driver intent</p></title>
  <aug>
    <au><snm>Sezer</snm><fnm>V</fnm></au>
    <au><snm>Bandyopadhyay</snm><fnm>T</fnm></au>
    <au><snm>Rus</snm><fnm>D</fnm></au>
    <au><snm>Frazzoli</snm><fnm>E</fnm></au>
    <au><snm>Hsu</snm><fnm>D</fnm></au>
  </aug>
  <source>2015 {IEEE}/{RSJ} {International} {Conference} on {Intelligent}
  {Robots} and {Systems} ({IROS})</source>
  <publisher>Hamburg, Germany: IEEE</publisher>
  <pubdate>2015</pubdate>
  <fpage>3578</fpage>
  <lpage>-3585</lpage>
</bibl>

<bibl id="B14">
  <title><p>Courtesy behavior for highly automated vehicles on highway
  interchanges</p></title>
  <aug>
    <au><snm>Menendez Romero</snm><fnm>C</fnm></au>
    <au><snm>Sezer</snm><fnm>M</fnm></au>
    <au><snm>Winkler</snm><fnm>F</fnm></au>
    <au><snm>Dornhege</snm><fnm>C</fnm></au>
    <au><snm>Burgard</snm><fnm>W</fnm></au>
  </aug>
  <source>2018 {IEEE} {Intelligent} {Vehicles} {Symposium} ({IV})</source>
  <publisher>Changshu: IEEE</publisher>
  <pubdate>2018</pubdate>
  <fpage>943</fpage>
  <lpage>-948</lpage>
</bibl>

<bibl id="B15">
  <title><p>Decision and behavior planning for a self-driving vehicle at
  unsignalized intersections</p></title>
  <aug>
    <au><snm>Wang</snm><fnm>WJ</fnm></au>
  </aug>
  <source>2020 {International} {Automatic} {Control} {Conference}
  ({CACS})</source>
  <publisher>Hsinchu, Taiwan: IEEE</publisher>
  <pubdate>2020</pubdate>
  <fpage>1</fpage>
  <lpage>-6</lpage>
</bibl>

<bibl id="B16">
  <title><p>Interaction-aware motion prediction for autonomous driving: A
  multiple model kalman filtering scheme</p></title>
  <aug>
    <au><snm>Lefkopoulos</snm><fnm>V</fnm></au>
    <au><snm>Menner</snm><fnm>M</fnm></au>
    <au><snm>Domahidi</snm><fnm>A</fnm></au>
    <au><snm>Zeilinger</snm><fnm>MN</fnm></au>
  </aug>
  <source>IEEE Robotics and Automation Letters</source>
  <pubdate>2021</pubdate>
  <volume>6</volume>
  <issue>1</issue>
  <fpage>80</fpage>
  <lpage>-87</lpage>
</bibl>

<bibl id="B17">
  <title><p>Design of longitudinal control for autonomous vehicles based on
  interactive intention inference of surrounding vehicle behavior using long
  short-term memory</p></title>
  <aug>
    <au><snm>Yoon</snm><fnm>Y</fnm></au>
    <au><snm>Yi</snm><fnm>K</fnm></au>
  </aug>
  <source>2021 {IEEE} {International} {Intelligent} {Transportation} {Systems}
  {Conference} ({ITSC})</source>
  <publisher>Indianapolis, IN, USA: IEEE</publisher>
  <pubdate>2021</pubdate>
  <fpage>196</fpage>
  <lpage>-203</lpage>
</bibl>

<bibl id="B18">
  <title><p>Multiple contextual cues integrated trajectory prediction for
  autonomous driving</p></title>
  <aug>
    <au><snm>Wang</snm><fnm>L</fnm></au>
    <au><snm>Wu</snm><fnm>T</fnm></au>
    <au><snm>Fu</snm><fnm>H</fnm></au>
    <au><snm>Xiao</snm><fnm>L</fnm></au>
    <au><snm>Wang</snm><fnm>Z</fnm></au>
    <au><snm>Dai</snm><fnm>B</fnm></au>
  </aug>
  <source>IEEE Robotics and Automation Letters</source>
  <pubdate>2021</pubdate>
  <volume>6</volume>
  <issue>4</issue>
  <fpage>6844</fpage>
  <lpage>-6851</lpage>
</bibl>

<bibl id="B19">
  <title><p>Vehicle motion prediction at intersections based on the turning
  intention and prior trajectories model</p></title>
  <aug>
    <au><snm>Zhang</snm><fnm>T</fnm></au>
    <au><snm>Song</snm><fnm>W</fnm></au>
    <au><snm>Fu</snm><fnm>M</fnm></au>
    <au><snm>Yang</snm><fnm>Y</fnm></au>
    <au><snm>Wang</snm><fnm>M</fnm></au>
  </aug>
  <source>IEEE/CAA Journal of Automatica Sinica</source>
  <pubdate>2021</pubdate>
  <volume>8</volume>
  <issue>10</issue>
  <fpage>1657</fpage>
  <lpage>-1666</lpage>
</bibl>

<bibl id="B20">
  <title><p>Scenario understanding and motion prediction for autonomous
  vehicles - Review and comparison</p></title>
  <aug>
    <au><snm>Karle</snm><fnm>P</fnm></au>
    <au><snm>Geisslinger</snm><fnm>M</fnm></au>
    <au><snm>Betz</snm><fnm>J</fnm></au>
    <au><snm>Lienkamp</snm><fnm>M</fnm></au>
  </aug>
  <source>IEEE Transactions on Intelligent Transportation Systems</source>
  <pubdate>2022</pubdate>
  <fpage>1</fpage>
  <lpage>-21</lpage>
</bibl>

<bibl id="B21">
  <title><p>Game-theoretic modeling of multi-vehicle interactions at
  uncontrolled intersections</p></title>
  <aug>
    <au><snm>Li</snm><fnm>N</fnm></au>
    <au><snm>Yao</snm><fnm>Y</fnm></au>
    <au><snm>Kolmanovsky</snm><fnm>I</fnm></au>
    <au><snm>Atkins</snm><fnm>E</fnm></au>
    <au><snm>Girard</snm><fnm>AR</fnm></au>
  </aug>
  <source>IEEE Transactions on Intelligent Transportation Systems</source>
  <publisher>IEEE</publisher>
  <pubdate>2020</pubdate>
</bibl>

<bibl id="B22">
  <title><p>A game-theoretic reinforcement learning approach for adaptive
  interaction at intersections</p></title>
  <aug>
    <au><snm>Jin</snm><fnm>X</fnm></au>
    <au><snm>Li</snm><fnm>K</fnm></au>
    <au><snm>Jia</snm><fnm>QS</fnm></au>
    <au><snm>Xia</snm><fnm>H</fnm></au>
    <au><snm>Bai</snm><fnm>Y</fnm></au>
    <au><snm>Ren</snm><fnm>D</fnm></au>
  </aug>
  <source>2020 Chinese Automation Congress (CAC)</source>
  <pubdate>2020</pubdate>
  <fpage>4451</fpage>
  <lpage>-4456</lpage>
</bibl>

<bibl id="B23">
  <title><p>Game Theoretic Modeling and Decision Making for Connected Vehicle
  Interactions at Urban Intersections</p></title>
  <aug>
    <au><snm>Cai</snm><fnm>J</fnm></au>
    <au><snm>Hang</snm><fnm>P</fnm></au>
    <au><snm>Lv</snm><fnm>C</fnm></au>
  </aug>
  <source>2021 6th IEEE International Conference on Advanced Robotics and
  Mechatronics (ICARM)</source>
  <pubdate>2021</pubdate>
  <fpage>874</fpage>
  <lpage>-880</lpage>
</bibl>

<bibl id="B24">
  <title><p>GamePlan: Game-Theoretic Multi-Agent Planning with Human Drivers at
  Intersections, Roundabouts, and Merging</p></title>
  <aug>
    <au><snm>Chandra</snm><fnm>R</fnm></au>
    <au><snm>Manocha</snm><fnm>D</fnm></au>
  </aug>
  <source>IEEE Robotics and Automation Letters</source>
  <publisher>IEEE</publisher>
  <pubdate>2022</pubdate>
</bibl>

<bibl id="B25">
  <title><p>Research on Autonomous Driving Decision Algorithm Considering
  Social Compatibility</p></title>
  <aug>
    <au><snm>Chen</snm><fnm>W</fnm></au>
  </aug>
  <source>PhD thesis</source>
  <publisher>Zhejiang University</publisher>
  <pubdate>2021</pubdate>
</bibl>

<bibl id="B26">
  <title><p>Socially-compatible behavior design of autonomous vehicles with
  verification on real human data</p></title>
  <aug>
    <au><snm>Wang</snm><fnm>L</fnm></au>
    <au><snm>Sun</snm><fnm>L</fnm></au>
    <au><snm>Tomizuka</snm><fnm>M</fnm></au>
    <au><snm>Zhan</snm><fnm>W</fnm></au>
  </aug>
  <source>IEEE Robotics and Automation Letters</source>
  <pubdate>2021</pubdate>
  <volume>6</volume>
  <issue>2</issue>
  <fpage>3421</fpage>
  <lpage>-3428</lpage>
</bibl>

<bibl id="B27">
  <title><p>Human-like driving decision at unsignalized intersections based on
  game theory</p></title>
  <aug>
    <au><snm>Li</snm><fnm>D</fnm></au>
    <au><snm>Liu</snm><fnm>G</fnm></au>
    <au><snm>Xiao</snm><fnm>B</fnm></au>
  </aug>
  <source>Proceedings of the Institution of Mechanical Engineers, Part D:
  Journal of Automobile Engineering</source>
  <pubdate>2022</pubdate>
</bibl>

<bibl id="B28">
  <title><p>Two-lane two-way overtaking decision model with driving style
  awareness based on a game-theoretic framework</p></title>
  <aug>
    <au><snm>Li</snm><fnm>D</fnm></au>
    <au><snm>Hao</snm><fnm>P</fnm></au>
  </aug>
  <source>Transportmetrica A: Transport Science</source>
  <pubdate>2022</pubdate>
</bibl>

<bibl id="B29">
  <title><p>Forming strategic alliances: The role of social
  compatibility</p></title>
  <aug>
    <au><snm>Ladegård</snm><fnm>G</fnm></au>
  </aug>
  <source>PhD thesis</source>
  <publisher>Norwegian School of Economics and Business
  Administration</publisher>
  <pubdate>1997</pubdate>
</bibl>

<bibl id="B30">
  <title><p>Personal space of traffic accident video</p></title>
  <aug>
    <au><cnm>{Traffic accident video}</cnm></au>
  </aug>
  <pubdate>2020</pubdate>
  <url>https://www.acfun.cn/u/4075269 (accessed~ Feb. 08, 2021)</url>
</bibl>

<bibl id="B31">
  <title><p>Methods of multidisciplinary in-depth analyses of road traffic
  accidents</p></title>
  <aug>
    <au><snm>Larsen</snm><fnm>L</fnm></au>
  </aug>
  <source>Journal of Hazardous Materials</source>
  <pubdate>2004</pubdate>
  <volume>111</volume>
  <issue>1</issue>
  <fpage>115</fpage>
  <lpage>122</lpage>
  <note>A Selection of Papers from the JRC/ESReDA Seminar on Safety
  Investigation Accidents, Petten, The Netherlands, 12-13 May, 2003</note>
</bibl>

<bibl id="B32">
  <title><p>DADA-2000: Can driving accident be predicted by driver attention?
  Analyzed by a benchmark</p></title>
  <aug>
    <au><snm>Fang</snm><fnm>J</fnm></au>
    <au><snm>Yan</snm><fnm>D</fnm></au>
    <au><snm>Qiao</snm><fnm>J</fnm></au>
    <au><snm>Xue</snm><fnm>J</fnm></au>
    <au><snm>Wang</snm><fnm>H</fnm></au>
    <au><snm>Li</snm><fnm>S</fnm></au>
  </aug>
  <source>2019 IEEE Intelligent Transportation Systems Conference
  (ITSC)</source>
  <publisher>Auckland, NZ</publisher>
  <pubdate>2019</pubdate>
  <fpage>4303</fpage>
  <lpage>-4309</lpage>
</bibl>

<bibl id="B33">
  <title><p>On a formal model of safe and scalable self-driving
  cars</p></title>
  <aug>
    <au><snm>Shalev Shwartz</snm><fnm>S</fnm></au>
    <au><snm>Shammah</snm><fnm>S</fnm></au>
    <au><snm>Shashua</snm><fnm>A</fnm></au>
  </aug>
  <publisher>arXiv</publisher>
  <pubdate>2018</pubdate>
</bibl>

<bibl id="B34">
  <title><p>A research on the collision avoidance strategy for autonomous
  emergency braking system</p></title>
  <aug>
    <au><snm>Li</snm><fnm>L</fnm></au>
    <au><snm>Zhu</snm><fnm>X</fnm></au>
    <au><snm>Dong</snm><fnm>X</fnm></au>
    <au><snm>Ma</snm><fnm>Z</fnm></au>
  </aug>
  <source>Automotive Engineering</source>
  <pubdate>2015</pubdate>
  <volume>37</volume>
  <issue>2</issue>
  <fpage>168</fpage>
  <lpage>-174</lpage>
</bibl>

<bibl id="B35">
  <title><p>Emotion recognition and application based on physiological
  signals</p></title>
  <aug>
    <au><snm>Lin</snm><fnm>W</fnm></au>
  </aug>
  <source>PhD thesis</source>
  <publisher>Zhejiang University</publisher>
  <pubdate>2019</pubdate>
</bibl>

<bibl id="B36">
  <title><p>Driving experience, personality, and skill and safety-motive
  dimensions in drivers’self-assessments</p></title>
  <aug>
    <au><snm>Lajunen</snm><fnm>T</fnm></au>
    <au><snm>Summala</snm><fnm>H</fnm></au>
  </aug>
  <source>Personality and Individual Differences</source>
  <pubdate>1995</pubdate>
  <volume>19</volume>
  <issue>3</issue>
  <fpage>307</fpage>
  <lpage>-318</lpage>
</bibl>

<bibl id="B37">
  <title><p>Analysis on lane changing trajectory under different driving style
  and design on assistant lane changing system</p></title>
  <aug>
    <au><snm>Liu</snm><fnm>J</fnm></au>
  </aug>
  <source>PhD thesis</source>
  <publisher>Changsha University of Science \& Technology</publisher>
  <pubdate>2015</pubdate>
</bibl>

<bibl id="B38">
  <title><p>Study on discretionary lane-changing behavior on urban
  streets</p></title>
  <aug>
    <au><snm>Sun</snm><fnm>Y</fnm></au>
  </aug>
  <source>PhD thesis</source>
  <publisher>Dalian University of Technology</publisher>
  <pubdate>2017</pubdate>
</bibl>

<bibl id="B39">
  <title><p>The research of the {EEG} frequency power features in three basic
  emotions</p></title>
  <aug>
    <au><snm>Cao</snm><fnm>K</fnm></au>
  </aug>
  <source>PhD thesis</source>
  <publisher>Tianjin Medical University</publisher>
  <pubdate>2019</pubdate>
</bibl>

</refgrp>
} 

\end{backmatter}
\end{document}